\documentclass[12pt,prd,nopacs,tightenlines,nofootinbib]{revtex4}
\usepackage{bm}
\usepackage{graphics}
\usepackage{rotating}
\usepackage{epsfig}
\usepackage{multirow}
\begin{document}
\title{
Relativistic description of the semileptonic decays of bottom mesons}
\author{R. N. Faustov $^1$}\email{faustov@ccas.ru}
\author{V. O. Galkin $^1$}\email{galkin@ccas.ru}
\author{Xian-Wei Kang $^{2, 3}$}\email{xwkang@bnu.edu.cn}

\affiliation{$^1$ Federal Research Center ``Computer Science and Control'', Russian Academy of Sciences,  Vavilov Street 40, 119333 Moscow, Russia\\
 $^2$ Key Laboratory of Beam Technology of the Ministry of Education,
College of Nuclear Science and Technology, Beijing Normal University, Beijing 100875, China\\
$^3$ Institute of Radiation Technology, Beijing Academy
of Science and Technology, Beijing 100875, China}

\begin{abstract}
The form factors of the semileptonic $B$, $B_s$ and $B_c$ meson decays
are calculated in the framework of the 
relativistic quark model based on the quasipotential approach in QCD. They
are expressed through the overlap integrals of the meson wave
function. All relativistic effects are consistently taken into
account. The momentum transfer $q^2$ behavior of form factors is
determined in the whole accessible kinematical range.  We
do not use any extrapolations,  heavy quark $1/m_Q$ expansion or model
assumptions about the shape of form factors. Convenient analytic
expressions of the form factors are given, which very accurately
reproduce the numerical results of our calculation. On the basis of these form factors and
helicity formalism, the differential and total branching fractions of
various semileptonic decays of bottom meson are calculated.   The mean values of the forward-backward asymmetry
$\langle A_{FB}\rangle$, lepton-side convexity parameter  $\langle
C^\ell_{F}\rangle$, longitudinal  $\langle P^\ell_{L}\rangle$ and
transverse $\langle P^\ell_{T}\rangle$ polarization of the charged
lepton, and the longitudinal
polarization fraction $\langle F_{L}\rangle$ for final-state vector meson are also evaluated. We
present a detailed comparison of the obtained predictions with the
calculations based on  the covariant light-front quark model and
confront them to available lattice QCD and experimental data. It is
found that although both models predict close values of the total
branching fractions, the differential distributions and forward-backward asymmetry and
polarization parameters differ significantly, especially for the
heavy-to-light semileptonic decays. We identify observables which
measurement can help to discriminate between models.
\end{abstract}

\maketitle

\section{Introduction}

The semileptonic decays are the main source of
the determination of the Cabibbo-Kobayashi-Maskawa (CKM) matrix
elements which are the fundamental parameters of the Standard
Model (SM).  The important feature of SM is the
universality of the electroweak coupling to all the three generations of
fermions, which leads to a lepton flavor universality. As a result the
lepton flavor symmetry between semileptonic decay rates involving
different lepton flavors arise when  the charged
lepton mass contributions to decay amplitudes and phase space are
taken into account. Thus
investigation of semileptonic decays can be used to test SM. Any
deviations from the CKM matrix unitarity constraints  and lepton
flavor universality will be a signal of the so-called new
physics. There is a longstanding tension between the values of
$|V_{cb}|$ and $|V_{ub}|$ extracted form experimental data \cite{pdg} on
exclusive and inclusive semileptonic decays (for a recent review see
Ref.~\cite{sdr}). Some hints of the possible lepton flavor violation, due to
the anomalously high rates for semileptonic $b\to c\tau\nu_\tau$
compared to SM predictions, were also reported (see,
e.g., Ref.~\cite{bsrw} for a recent review). However, the most recent experimental data tend to
decrease these deviations.

The experimental and theoretical studies of  semileptonic decays of
bottom mesons, which are governed by $b\to c$ and $b\to u$ quark transitions,
 are important for the determination of the CKM matrix
elements $|V_{cb}|$ and  $|V_{ub}|$. The semileptonic decays of $B_c$
meson to $B_s$ and $B$ mesons proceed through $c\to s$ and $c\to d$ quark
transitions, and thus are proportional to the CKM matrix elements $|V_{cs}|$
and  $|V_{cd}|$. The main theoretical difficulty in considering exclusive
semileptonic decays of bottom mesons is related to the calculation of
the form factors, which parameterize the hadronic matrix elements,
since the lepton part can be easily calculated by the standard methods. It
is important to reliably determine the momentum transfer squared $q^2$
behavior of these form factors in the whole accessible kinematical
range, since differential distributions are very sensitive to it. This
is especially important for heavy-to-light decays since they have a
very broad $q^2$ range. Most of the theoretical approaches determine
these form factors at some particular kinematical points or in the
limited $q^2$ range and then extrapolate them using some model
parameterizations. For example, light cone sum rules \cite{lcsr} calculate form
factors at the maximum recoil point of the final meson $q^2=0$ (the
small electron mass is neglected), while
in lattice QCD calculations  the high $q^2$ region is
investigated  (for a review see, e.g., Ref.~\cite{lqcdr} and
references therein). And also in most considerations of heavy-to-heavy 
($B\to D^{(*)}$) semileptonic decays the heavy quark $1/m_Q$ expansion
is employed. Some very recent works on the semileptonic bottom meson decay are in e.g.,
\cite{Heger:2021gxt,Zhou:2022fyb,Xing:2022sor}, where it is interesting that in Ref.~\cite{Heger:2021gxt}
both the space- and time-like momentum transfer regions are considered.

In the present paper we use the relativistic quark model
based on the quasipotential approach in QCD for the calculations of the
bottom meson transition form factors in the whole accessible kinematical
range. The comprehensive account of the relativistic effects allows us to
achieve this goal without any expansions and extrapolations. On this
basis we calculated semileptonic decay branching fractions of bottom
mesons as well as differential distributions, foward-backward (FB) asymmetry and
polarization parameters. Similar approach was recently used by us for
consideration of the $D$ meson semileptonic decays \cite{fgk}. We
present also the detailed comparison of our predictions with the results of
the other popular quark model --- the covariant light-front quark
model (CLFQM) \cite{v,zkgdlw} and identify observables which measurement can help to
discriminate between models.  Model predictions are also compared with
available data from the experiment and lattice QCD.

\section{Relativistic quark model }
\label{sec:rqm}

We use the relativistic quark model (RQM) based on the quasipotential
approach to calculate the form factors parameterizing the matrix
elements of the weak $B$, $B_s$ and $B_c$ meson decays.
In this model a meson $M$ is considered as a quark-antiquark bound state
described by the wave function $\Psi_M$, which satisfies the
three-dimensional relativistically invariant
Schr\"odinger-type quasipotential equation \cite{efg}
\begin{equation}
\label{quasipot}
{\left(\frac{b^2(M)}{2\mu_{R}}-\frac{{\bf
p}^2}{2\mu_{R}}\right)\Psi_{M}({\bf p})} =\int\frac{d^3 q}{(2\pi)^3}
 V({\bf p,q};M)\Psi_{M}({\bf q}),
\end{equation}
where
\begin{equation}
{b^2(M) }
=\frac{[M^2-(m_1+m_2)^2][M^2-(m_1-m_2)^2]}{4M^2}
\end{equation}
and
\begin{equation}
  \mu_{R}=\frac{M^4-(m^2_1-m^2_2)^2}{4M^3}
\end{equation}
are  the  relative momentum squared on mass
shell in the center of mass system and  the relativistic reduced mass,
respectively.
Here $M$ is the meson mass, $m_{1,2}$ are the quark masses,
and ${\bf p}$ is their relative momentum.

The kernel of this equation $V({\bf p,q};M)$ is the
quark-antiquark interaction quasipotential. In RQM it is constructed \cite{efg} with the
help of the
QCD-motivated off-mass-shell scattering amplitude projected on the
positive energy states. The effective quark-antiquark interaction is
chosen as the sum of the one-gluon exchange potential, which dominates
at short distances, and the long range confining interaction linearly
increasing with a quark separation, which dominates at large distances. The Lorentz-structure of the
confining interaction is assumed to be the mixture of the vector and
scalar terms with the mixing coefficient $\epsilon$. It is also
assumed that the vertex of the vector confining interaction
$\Gamma_\mu({\bf k})$ with ${\bf k=p-q}$ contains both Dirac and Pauli terms
\begin{equation}
\label{kappa}
\Gamma_{\mu}({\bf k})=\gamma_{\mu}+
\frac{i\kappa}{2m}\sigma_{\mu\nu}k^{\nu},
\end{equation}
thus, introducing the long-range anomalous
chromomagnetic quark moment $\kappa$.   In the nonrelativistic limit
this quasipotential reduces to the standard Cornell potential
\begin{equation}
V(r)=-\frac43\frac{\alpha_s}{r}+Ar+B,
\end{equation}
where $\alpha_s$ is the QCD coupling constant. Therefore, our
quasipotential can be considered as its relativistic generalization,
which takes into account both spin-independent and spin-dependent
contributions. Its explicit form can be found in Ref.~\cite{efglregge}.

All parameters of the model were fixed from previous considerations of
meson properties \cite{efg,efglregge,efghlregge,efghregge}. The constituent quark masses $m_u=m_d=0.33$~GeV,
$m_s=0.5$~GeV, $m_c=1.55$~GeV and $m_b=4.88$~GeV and parameters of
confining potential $A=0.18$~GeV$^2$ and $B=-0.30$~GeV have values
standard for the quark models. Our extra parameters: the mixing
coefficient of vector and scalar confining potentials is $\epsilon=-1$
and the long-range anomalous chromomagnetic moment is $\kappa=-1$. For
the chosen value of $\kappa$, the long range chromomagnetic
interaction, which is proportional to $(1+\kappa)$, vanishes in accord
with the flux tube model.

The spectroscopy of the heavy \cite{efghregge}, heavy-light \cite{efghlregge} and
light \cite{efglregge} mesons was considered previously. The quasipotential
equation (\ref{quasipot}) with the complete relativistic quasipotential was
solved numerically. Masses of ground and excited mesons were found in
good agreement with available experimental data. The numerical wave
functions of the meson states were obtained. Here we apply these wave
functions for the calculations of the weak decay form factors of
$B$, $B_s$ and $B_c$ mesons.

In the quasipotential approach the matrix element of the weak current $J^W_\mu=\bar
q\gamma_\mu(1-\gamma_5)b$ ($q=u, c$) between initial ($B=B, B_s, B_c$) and final
($F=D^{(*)}, D_s^{(*)}, \pi, \rho, \eta, \eta', \omega, K^{(*)}, \eta_c, J/\psi$)
meson states is expressed as \cite{efgsml,fgsml,fgbs}
\begin{equation}\label{mel}
\langle F(p_{F}) \vert J^W_\mu \vert B(p_{B})\rangle
=\int \frac{d^3p\, d^3q}{(2\pi )^6} \bar \Psi_{{F}\,{\bf p}_F}({\bf
p})\Gamma _\mu ({\bf p},{\bf q})\Psi_{B\,{\bf p}_{B}}({\bf q}),
\end{equation}
where $p_B$ and $p_F$ are the initial and final meson momenta,
respectively. The wave functions $\Psi_{M\,{\bf p}_M}$ of the
initial ($M=B$) and final ($M=F$) mesons are projected on the positive energy states
and boosted to the moving reference frame with the three-momentum
${\bf p}_M$. The wave function of the moving meson $\Psi_{M\,{\bf  p}_M}$ is connected with the wave function in the rest reference
frame $\Psi_{M\,{\bf 0}}$ by the transformation \cite{efgsml}
 \begin{equation}
\label{wig}
\Psi_{M\,{\bf p}_M}({\bf p})=D_q^{1/2}(R_{L_{{\bf p}_M}}^W)
  D_{\bar  q_s}^{1/2}(R_{L_{{\bf p}_M}}^W)\Psi_{M\,{\bf 0}}({\bf p}),
\end{equation}
where $q$ and $\bar  q_s$ are the active quark and spectator
antiquark, respectively. $R^W$ is the Wigner rotation, $L_{{\bf p}_M}$ is the Lorentz boost
from the meson rest frame to a moving one, and $D^{1/2}(R)$ is
the spin rotation matrix
\[
{1 \ \ \,0\choose 0 \ \ \,1}D^{1/2}_{q}(R^W_{L_{{\bf p}_M}})=
S^{-1}({\bf p}_{q})S({\bf p}_M)S({\bf p}),\]
with
\[
S({\bf p})=\sqrt{\frac{\epsilon({\bf p})+m}{2m}}\left(1+\frac{\bm{\alpha}\cdot{\bf p}}
{\epsilon({\bf p})+m}\right),
\]
with $\bm{\alpha} =\gamma_0\bm{\gamma}$ being the product of the Dirac matrices, and $m$
is the corresponding quark or meson mass.

The vertex function of the $W$-boson interaction
consists of two terms $\Gamma _\mu ({\bf p},{\bf q})=\Gamma^{(1)}
_\mu ({\bf p},{\bf q})+\Gamma^{(2)} _\mu ({\bf p},{\bf q})$. The first
term corresponds to the impulse approximation, where there is no
interaction between the active quark and spectator antiquark. Thus it
contains the $\delta({\bf p}_s-{\bf q}_s)$ function which is
responsible for the momentum conservation on the spectator $\bar q_s$ line.  The
second term is the consequence of the projection on the positive
energy states and takes into account interaction of the negative
energy part of the active quark with the spectator antiquark,
so-called $Z$-diagrams. Thus, this term includes the $q\bar q$
interaction quasipotentials $V({\bf p,q};M)$ and the negative-energy
parts of the quark propagators. The corresponding
diagrams and explicit expressions for the vertex functions  $\Gamma^{(1),(2)}
_\mu ({\bf p},{\bf q})$ are given in Ref.~\cite{efgsml}.

It is convenient to express the weak decay matrix element, Eq.~(\ref{mel}),
in the form of the overlap integral of the meson wave functions. It can be
easily done for the leading contribution $\Gamma^{(1)}_\mu ({\bf p},{\bf q})$,
since it contains the $\delta$-function, which allows one to take one of
the integrations in Eq.~(\ref{mel}). The second contribution
$\Gamma^{(2)}_\mu ({\bf p},{\bf q})$ is more complicated. Instead of
the $\delta$-function it contains the quasipotential $V({\bf p,q};M)$
with the nontrivial Lorentz-structure. However, it is possible to use the
quasipotential equation to take off one of the integrations in
Eq.~(\ref{mel}) and thus to get the desired form of the matrix elements
(for details see Ref.~\cite{efgsml}). As a result this
contribution is proportional to the ratio of the meson binding energy
to the energy of the active quark. Therefore, it is indeed the
subleading contribution. It turns out to be rather small for all
heavy-to-heavy meson weak decays.  For the heavy-to-light decays its
contribution is suppressed for large recoils of the final meson due to
the large energy of the final active quark  but
becomes important at small recoils.

\section{Form factors of semileptonic decays}
\label{sec:ff}

The hadronic  matrix element of the weak current $J^W_\mu$ between meson
states is usually parameterized by the following set of invariant form factors.
\begin{itemize}
\item  For $B$ transition to the pseudoscalar $P$ meson (where $B$ symbolically denotes either $B,B_s$ or $B_c$))
\begin{eqnarray}
  \label{eq:psff}
 \langle P(p_P)|\bar q \gamma^\mu b|B(p_B)\rangle
 & =&f_+(q^2)\left[p_B^\mu+ p_P^\mu-
\frac{M_{B}^2-M_{P}^2}{q^2}\ q^\mu\right]+
  f_0(q^2)\frac{M_{B}^2-M_{P}^2}{q^2}\ q^\mu,\cr\cr
 \langle P(p_{P})|\bar q \gamma^\mu\gamma_5 b|B(p_{B})\rangle
  &=&0,
\end{eqnarray}
\item  For $B$ transition to the vector $V$  meson
\begin{eqnarray}
  \label{eq:vff}
 \langle V(p_V)|\bar q \gamma^\mu b|B(p_B)\rangle&=
  &\frac{2iV(q^2)}{M_{B}+M_{V}} \epsilon^{\mu\nu\rho\sigma}\epsilon^*_\nu
  p_{B\rho} p_{V\sigma},\cr \cr
\label{eq:vff2}
\langle V(p_{V})|\bar q \gamma^\mu\gamma_5 b|B(p_{B})\rangle&=&2M_{V}
A_0(q^2)\frac{\epsilon^*\cdot q}{q^2}\ q^\mu
 +(M_{B}+M_{V})A_1(q^2)\left(\epsilon^{*\mu}-\frac{\epsilon^*\cdot
    q}{q^2}\ q^\mu\right)\cr\cr
&&-A_2(q^2)\frac{\epsilon^*\cdot q}{M_{B}+M_{V}}\left[p_{B}^\mu+
  p_{V}^\mu-\frac{M_{B}^2-M_{V}^2}{q^2}\ q^\mu\right],
\end{eqnarray}
\end{itemize}
here $q=p_B-p_F$ ($F=P,V$).
At the maximum recoil point ($q^2=0$) these form
factors satisfy the following conditions:
\[f_+(0)=f_0(0),\]
\[A_0(0)=\frac{M_{B}+M_{V}}{2M_{V}}A_1(0)
  -\frac{M_{B}-M_{V}}{2M_{V}}A_2(0).\]

These form factors are calculated in RQM  with the systematic consideration
of all relativistic effects as described in the previous
section. Relativistic transformations of the meson wave functions from
the rest reference frame to the moving one, cf. Eq.~(\ref{wig}), as well as
relativistic contributions from the intermediate negative-energy
states are consistently taken into account.  The explicit expressions
for these form factors in terms of the overlap integrals of initial
and final meson wave functions can be found in Ref.~\cite{efgbc}. It
was shown previously \cite{efgsml,fghq}, that applying the heavy quark expansion to these
expressions for the case of heavy-to-heavy meson transitions, it is possible to
reproduce the model independent expressions of heavy quark effective
theory at leading, first and second order in $1/m_Q$ and obtain
expressions for the corresponding Isgur-Wise functions.   In
the present paper we do not use the heavy quark expansion and
calculate all form factors nonperturbatively employing relativistic
meson wave functions. Such approach allows us to reliably determine
the momentum transfer squared ($q^2$) dependence of these form factors
in the whole accessible kinematic range  \cite{efgsml,fgsml,fgbs}. The obtained numerical results  can
be approximated with high accuracy by the following expressions:
\begin{itemize}
\item  for the form factors $f_+(q^2),V(q^2),A_0(q^2)$
\begin{equation}
  \label{fitfv}
  F(q^2)=\frac{F(0)}{\displaystyle\left(1-\frac{q^2}{ M^2}\right)
    \left(1-\sigma_1
      \frac{q^2}{M_{1}^2}+ \sigma_2\frac{q^4}{M_{1}^4}\right)},
\end{equation}

\item for the form factors $f_0(q^2), A_1(q^2),A_2(q^2)$
\begin{equation}
  \label{fita12}
  F(q^2)=\frac{F(0)}{\displaystyle \left(1-\sigma_1
      \frac{q^2}{M_{1}^2}+ \sigma_2\frac{q^4}{M_{1}^4}\right)},
\end{equation}
\end{itemize}
where the masses $M$ and $M_1$ are given in Table~\ref{mpf}. For the
decays governed by the CKM favored $b\to c$ transitions, for pole
masses $M$ we use the
masses of the intermediate vector $B_c^*$ mesons for the
form factors $f_+(q^2),V(q^2)$ and of the pseudoscalar $B_c$ for the form
factor $A_0(q^2)$. While for the decays governed by the CKM
suppressed $b\to u$ transitions, masses of the intermediate $B^*$ and $B$
mesons are used, respectively. For $B_c\to B_s$
decays governed by the CKM favored $c\to s$ transitions and $B_c\to B$
decays governed by the CKM
suppressed $c\to d$ transitions, for pole masses $M$ we use the
respective masses of vector
$D_s^*$ and $D^*$ mesons and pseudoscalar
$D_s$ and $D$ mesons.  The values of form
factors $F(0)$, $F(q^2_{\rm max})$ and fitted parameters
$\sigma_{1,2}$ for the weak decays of $B$  and $B_s$  mesons
are given in Table~\ref{ff1} and of the $B_c$ meson in Table~\ref{ff2}. We estimate
the uncertainties of the calculated form factors to be less than 5\%.

\begin{table}
\caption{Masses in parameterizations of the weak decay form factors of $B$, $B_s$ and  $B_c$ mesons, cf. Eqs.~(\ref{fitfv}) and
  (\ref{fita12}). }
\label{mpf}
\begin{ruledtabular}
  \begin{tabular}{ccccc}
Quark transition&Decay& $M_1$ (GeV)&\multicolumn{2}{c}{$M$ (GeV)}\\
    \cline{4-5}
                &&&$f_+(q^2),V(q^2)$& $A_0(q^2)$\\ \hline
    $b\to c$&$B\to D(D^*)$&6.332&6.332&6.277\\
                &$B_s\to D_s(D_s^*)$&\\
                &$B_c\to \eta_c(J/\psi)$&   \\                 \hline
    $b\to u$&$B\to \pi(\rho)$&5.325&5.325&5.280\\
    &$B\to \eta^{(')}(\omega)$&\\
                &$B_s\to K(K^*)$&\\
                &$B_c\to D(D^*)$&   \\                 \hline
    $c\to s$&$B_c\to B_s(B_s^*)$&6.332&2.112&1.968\\ \hline
    $c\to d$&$B_c\to B(B^*)$&6.332&2.010&1.870\\
\end{tabular}
\end{ruledtabular}
\end{table}

\begin{table}
\caption{Form factors of the weak $B$ and $B_s$  meson transitions in RQM. }
\label{ff1}
\begin{ruledtabular}
\begin{tabular}{cccccc}
Transition& Form factor & $F(0)$& $F(q^2_{\rm max})$& $\sigma_1$ & $\sigma_2$ \\
\hline
  $B\to D$         &$f_+$ &$0.696$ & $1.24$ & 1.288 & $1.943$\\
   & $f_0$  &$0.696$ & $0.82$ & 0.731& $0.736$ \\ \hline
$B\to D^*$   &$V$&$0.915$& $1.38$& $0.647$ &$1.100$\\
     &$A_0$&$0.814$&$1.21$&$0.645$&$1.300$\\
     &$A_1$&$0.730$& $0.83$& $0.571$ &$0.457$\\
     &$A_2$&$0.627$& $1.02$& $0.719$ &$-2.690$\\ \hline
  $B\to \pi$   &$f_+$ &$0.217$ & $10.9$ & 0.378 & $-0.410$\\
     & $f_0$  &$0.217$ & $1.32$ & $-0.501$& $-1.50$ \\ \hline
$B\to \rho$   &$V$&$0.295$& $2.80$& $0.875$ &$0$\\
     &$A_0$&$0.231$&$2.19$&$0.796$&$-0.055$\\
     &$A_1$&$0.269$& $0.439$& $0.540$ &$0$\\
     &$A_2$&$0.282$& $1.92$& $1.34$ &$0.210$\\ \hline
$B\to \eta$   &$f_+$ &$0.194$ & $2.75$ & 0.181 & $-0.835$\\
     & $f_0$  &$0.194$ & $0.52$ & $0.458$& $-0.424$ \\ \hline
$B\to \eta'$   &$f_+$ &$0.187$ & $1.34$ & $0.637$ & $-0.396$\\
   & $f_0$  &$0.187$ & $0.32$ &  $0.996$& $0.565$ \\ \hline
$B\to \omega$   &$V$&$0.263$& $2.47$& $0.894$ &$0.021$\\
     &$A_0$&$0.244$&$1.96$&$0.826$&$0.084$\\
     &$A_1$&$0.232$& $0.46$& $0.676$ &$-0.032$\\
     &$A_2$&$0.229$& $0.96$& $1.453$ &$0.539$\\ \hline
  $B_s\to D_s$         &$f_+$ &$0.663$ & $1.23$ & 1.375 & $1.877$\\
   & $f_0$  &$0.663$ & $0.87$ & 1.018& $0.705$ \\ \hline
$B_s\to D_s^*$   &$V$&$0.925$& $1.48$& $0.965$ &$1.534$\\
     &$A_0$&$0.625$&$1.03$&$0.457$&$-0.710$\\
     &$A_1$&$0.668$& $0.82$& $0.913$ &$0.766$\\
     &$A_2$&$0.723$& $1.09$& $1.349$ &$0.230$\\ \hline
   $B_s\to K$         &$f_+$ &$0.284$ & $5.42$ & $-0.370$ & $-1.41$\\
   & $f_0$  &$0.284$ & $0.459$ & $-0.072$& $-0.651$ \\ \hline
$B_s\to K^*$   &$V$&$0.291$& $3.06$& $-0.516$ &$-2.10$\\
     &$A_0$&$0.289$&$2.10$&$-0.383$&$-1.58$\\
     &$A_1$&$0.287$& $0.581$& $0$ &$-1.06$\\
     &$A_2$&$0.286$& $0.953$& $1.05$ &$0.074$\\
\end{tabular}
\end{ruledtabular}
\end{table}

  \begin{table}
\caption{Form factors of the weak $B_c$ meson transitions in RQM. }
\label{ff2}
\begin{ruledtabular}
\begin{tabular}{cccccc}
Transition& Form factor & $F(0)$& $F(q^2_{\rm max})$& $\sigma_1$ & $\sigma_2$ \\
\hline
   $B_c\to \eta_c$         &$f_+$ &$0.431$ & $1.09$ & 2.103 & $1.510$\\
   & $f_0$  &$0.431$ & $0.93$ & 2.448& $1.700$ \\ \hline
$B_c\to J/\psi$   &$V$&$0.513$& $1.32$& $1.815$ &$-0.320$\\
     &$A_0$&$0.360$&$0.93$&$2.335$&$1.721$\\
     &$A_1$&$0.459$& $0.89$& $2.415$ &$1.929$\\
     &$A_2$&$0.653$& $1.33$& $2.765$ &$2.939$\\ \hline
  $B_c\to D$         &$f_+$ &$0.081$ & $3.21$ & $2.167$ & $1.203$\\
   & $f_0$  &$0.081$ & $0.63$ & $2.455$& $1.729$ \\ \hline
  $B_c\to D^*$   &$V$&$0.125$& $3.11$& $2.247$ &$1.346$\\
     &$A_0$&$0.035$&$0.99$&$1.511$&$0.175$\\
     &$A_1$&$0.054$& $0.78$& $2.595$ &$1.784$\\
     &$A_2$&$0.071$& $1.25$& $2.800$ &$2.073$\\ \hline
  $B_c\to B_s$         &$f_+$ &$0.524$ & $0.990$ & $10.91$ & $-292.1$\\
   & $f_0$  &$0.524$ & $0.822$ & $19.47$& $92.36$ \\ \hline
  $B_c\to B_s^*$   &$V$&$2.08$& $4.78$& $18.88$ &$-372.8$\\
     &$A_0$&$0.431$&$0.736$&$12.23$&$-141.5$\\
     &$A_1$&$0.518$& $0.850$& $19.11$ &$-103.6$\\
     &$A_2$&$1.62$& $2.02$& $17.47$ &$366.5$\\ \hline
   $B_c\to B$         &$f_+$ &$0.394$ & $0.969$ & $4.970$ & $-550.2$\\
   & $f_0$  &$0.394$ & $0.680$ & $19.92$& $117.7$ \\ \hline
  $B_c\to B^*$   &$V$&$1.84$& $4.79$& $11.48$ &$-477.1$\\
     &$A_0$&$0.380$&$0.906$&$11.32$&$-349.7$\\
     &$A_1$&$0.457$& $0.822$& $19.55$ &$-2.685$\\
     &$A_2$&$1.33$& $1.92$& $22.82$ &$407.6$\\
\end{tabular}
\end{ruledtabular}
\end{table}

\section{Semileptonic decays of bottom mesons}
\label{sec:sd}

The differential distribution of the $B$ meson
semileptonic decay to the final ground state pseudoscalar or
vector meson ($F=P,V$) has the following expression in terms of the helicity components \cite{ikpsst,zkgdlw}:
\begin{eqnarray}
  \label{eq:dGamma}
  \frac{d\Gamma(B\to
  F\ell^+\nu_\ell)}{dq^2d(\cos\theta)}&=&\frac{G_F^2}{(2\pi)^3}
  |V_{q_1q_2}|^2\frac{\lambda^{1/2}q^2}{64M_{B}^3}\left(1-\frac{m_\ell^2}{q^2}\right)^2\Biggl[(1+\cos^2\theta)
{\cal H}_U+2\sin^2\theta{\cal H}_L+2\cos\theta{\cal H}_P\cr
&&+\frac{m_\ell^2}{q^2}(\sin^2\theta{\cal H}_U
   +2\cos^2\theta{\cal H}_L+ 2 {\cal H}_S-4\cos\theta {\cal H}_{SL})\Bigr],
\end{eqnarray}
where $V_{q_1q_2}$ is the corresponding CKM
matrix element, $\lambda\equiv
\lambda(M_{B}^2,M_F^2,q^2)=M_{B}^4+M_F^4+q^4-2(M_{B}^2M_F^2+M_F^2q^2+M_{B}^2q^2)$,
$m_\ell$ is the lepton mass ($\ell=e,\mu,\tau$), and the polar angle $\theta$ is the angle
between the momentum of the charged lepton in the rest frame of the intermediate
$W$-boson and the direction opposite to the final $F$ meson momentum in the rest frame of the initial $B$ meson.  ${\cal
  H}_I$  ($I=U,L,P,S,SL$) are the bilinear combinations of the helicity components of the
hadronic tensor \cite{ikpsst,zkgdlw}:
\begin{equation}
  \label{eq:hh}
{\cal H}_U=|H_+|^2+|H_-|^2, \quad {\cal H}_L=|H_0|^2, \quad {\cal
   H}_P=|H_+|^2-|H_-|^2,\quad
    {\cal H}_S=|H_t|^2, \quad {\cal H}_{SL}=\Re(H_0H_t^\dag),
  \end{equation}
where the helicity amplitudes $H_i$ ($i=+,-,0,t$) are the following combinations of invariant form
factors considered in the previous section.
\begin{itemize}
\item For $B$ to the pseudoscalar $P$ meson transitions, the helicity
  amplitudes are given by
  \begin{equation}
  \label{eq:hap}
  H_\pm=0,\quad
H_0=\frac{\lambda^{1/2}}{\sqrt{q^2}}f_+(q^2),\quad
H_t=\frac1{\sqrt{q^2}}(M_{B}^2-M_{P}^2)f_0(q^2).
\end{equation}
\item For $B$ to the vector $V$ meson transitions, the helicity
  amplitudes are the following
  \begin{eqnarray}
  \label{eq:hav}
  H_\pm(q^2)&=&(M_{B}+M_{V})A_1(q^2)\mp \frac{\lambda^{1/2}}{M_{B}+M_{V}}V(q^2),\cr
  H_0(q^2)&=&\frac1{2M_{V}\sqrt{q^2}}\left[(M_{B}+M_{V})
(M_{B}^2-M_{V}^2-q^2)A_1(q^2)-\frac{\lambda}{M_{B}
+M_{V}}A_2(q^2)\right], \cr
    H_t&=&\frac{\lambda^{1/2}}{\sqrt{q^2}}A_0(q^2).
\end{eqnarray}
\end{itemize}

The expression (\ref{eq:dGamma}), normalized by the differential decay distribution integrated
over   $\cos\theta$
\begin{equation}
  \label{eq:dg}
d\Gamma/dq^2\equiv  \frac{d\Gamma(B\to
  F\ell^+\nu_\ell)}{dq^2}=\frac{G_F^2}{(2\pi)^3}
  |V_{q_1q_2}|^2\frac{\lambda^{1/2}q^2}{24M_{B}^3}\left(1-\frac{m_\ell^2}{q^2}\right)^2{\cal H}_{\rm total},
\end{equation}
can be rewritten as
\begin{equation}
  \label{eq:ndr}
\frac1{d\Gamma/dq^2} \frac{d\Gamma(B\to
  F\ell^+\nu_\ell)}{dq^2d(\cos\theta)} = \frac12\left[1-\frac13
  C^\ell_F(q^2)\right] +A_{FB}(q^2)\cos\theta+\frac12 C^\ell_F(q^2)\cos^2\theta.
\end{equation}
Here ${\cal H}_{\rm total}$ is the total helicity structure
\begin{equation}
  \label{eq:htot}
 {\cal H}_{\rm total}= ({\cal H}_U+{\cal
   H}_L)\left(1+\frac{m_\ell^2}{2q^2}\right) +\frac{3m_\ell^2}{2q^2}{\cal H}_S.
\end{equation}
In the differential decay distribution, Eq.~(\ref{eq:ndr}), $A_{FB}(q^2)$ is the FB asymmetry defined by
\begin{equation}
  \label{eq:afb}
  A_{FB}(q^2)=\frac34\frac{{\cal
      H}_P-2\frac{m_\ell^2}{q^2}{\cal H}_{SL}}{{\cal H}_{\rm total}},
\end{equation}
and  $C^\ell_F(q^2)$ is the lepton-side convexity parameter, which is the second derivative of
the distribution (\ref{eq:ndr}) over $\cos\theta$, given by
\begin{equation}
  \label{eq:clf}
   C^\ell_F(q^2)=\frac34\left(1-\frac{m_\ell^2}{q^2}\right)\frac{{\cal H}_U-2{\cal H}_L}{{\cal H}_{\rm total}}.
\end{equation}

The longitudinal polarization of the final charged lepton $\ell$ is
defined as the ratio of the longitudinally polarized decay
distribution to the unpolarized decay distribution, Eq .~(\ref{eq:dg}) \cite{ikpsst,zkgdlw}:
\begin{equation}
  \label{eq:ple}
  P_L^\ell(q^2)=\frac1{d\Gamma/dq^2} \frac{d\Gamma(s_L)}{dq^2}=\frac{({\cal H}_U+{\cal
   H}_L)\left(1-\frac{m_\ell^2}{2q^2}\right) -\frac{3m_\ell^2}{2q^2}{\cal H}_S}{{\cal H}_{\rm total}},
\end{equation}
and its transverse polarization is given by \cite{ikpsst,zkgdlw}
\begin{equation}
  \label{eq:pte}
  P_T^\ell(q^2)=\frac1{d\Gamma/dq^2} \frac{d\Gamma(s_T)}{dq^2}=-\frac{3\pi m_\ell}{8\sqrt{q^2}}\frac{{\cal H}_P+2{\cal H}_{SL}}{{\cal H}_{\rm total}},
\end{equation}
where $s_{L(T)}$ is the longitudinal (transverse) polarization vector
of the final lepton $\ell$,
which is parallel (perpendicular) to the lepton momentum.

For the $B$ decays to the vector $V$ meson, which then decays to two
pseudoscalar mesons $V\to P_1P_2$, the differential distribution in
the angle $\theta^*$, defined as the polar angle between the vector meson
$V$ momentum in the $B$ meson rest frame and the
pseudoscalar meson $P_1$ momentum in the rest frame of the vector meson $V$, is given by \cite{ikpsst,zkgdlw}
\begin{equation}
  \label{eq:mpdr}
\frac1{d\Gamma/dq^2} \frac{d\Gamma(B\to
  V(\to P_1P_2)\ell^+\nu_\ell)}{dq^2d(\cos\theta^*)} =
\frac34\left[2F_L(q^2)\cos^2\theta^* +F_T(q^2)\sin^2\theta^*\right].
\end{equation}
Here  the longitudinal polarization fraction
of the final vector meson has the following form in terms of the
helicity structures \cite{ikpsst,zkgdlw}:
\begin{equation}
  \label{eq:fl}
  F_L(q^2)=\frac{{\cal
   H}_L\left(1+\frac{m_\ell^2}{2q^2}\right) +\frac{3m_\ell^2}{2q^2}{\cal H}_S}{{\cal H}_{\rm total}},
\end{equation}
and its transverse polarization fraction $F_T(q^2)=1- F_L(q^2)$.

For $\bar B \to F \ell^-\bar \nu_\ell$ decays the charge of the lepton
is negative,  and thus expressions for the FB asymmetry,
leptonic longitudinal and transverse polarization change due to the
different sign in the leptonic tensor. They are the following
\begin{eqnarray}
  \label{eq:abarb}
  A_{FB}(q^2)&=&-\frac34\frac{{\cal
      H}_P+2\frac{m_\ell^2}{q^2}{\cal H}_{SL}}{{\cal H}_{\rm
                 total}},\\
  P_L^\ell(q^2)&=&-\frac{({\cal H}_U+{\cal
   H}_L)\left(1-\frac{m_\ell^2}{2q^2}\right)
                   -\frac{3m_\ell^2}{2q^2}{\cal H}_S}{{\cal H}_{\rm
                   total}},\\
  P_T^\ell(q^2)&=&-\frac{3\pi m_\ell}{8\sqrt{q^2}}\frac{{\cal H}_P-2{\cal H}_{SL}}{{\cal H}_{\rm total}}.
\end{eqnarray}
The expressions for $C^\ell_{F}(q^2)$ and $F_{L(T)}(q^2)$ do not change.

The mean values of the FB  asymmetry $\langle A_{FB}\rangle$ and
polarization  $\langle C^\ell_{F}\rangle$, $\langle
P^\ell_{L,T}\rangle$, $\langle
F_{L}\rangle$  parameters are
calculated by separately integrating their corresponding numerators and denominators
with inclusion of the common kinematical factor $\lambda^{1/2}q^2(1-m_\ell^2/q^2)^2$.

We give predictions for the branching fractions, FB asymmetry and
polarization parameters  of semileptonic $B, B_s$ and $B_c$ meson decays
calculated with the form factors of RQM  in
Tables~\ref{asB}-\ref{asBc}. The branching fractions are calculated
using the following values of CKM matrix elements: $|V_{cb}|=0.039$,
$|V_{ub}|=0.0040$, $|V_{cs}|=0.987$, $|V_{cd}|=0.221$. All meson and
lepton masses are taken from PDG \cite{pdg}. We roughly estimate the
uncertainties of our predictions for the branching fraction values
given in Tables~\ref{asB}-\ref{asBc}, which
arise from the model  to be about 10\%, and if we also include uncertainties
of the CKM matrix elements the estimate of the total uncertainty is
about 15\%.

\begin{table}
\caption{The branching ratios, FB asymmetry and polarization parameters
  of semileptonic $B$ decays in RQM.}
\label{asB}
\begin{ruledtabular}
  \begin{tabular}{ccccccc}
Decay&Br&$\langle A_{FB}\rangle$  & $\langle C^\ell_{F}\rangle$&
$\langle P^\ell_{L}\rangle$  &$\langle P^\ell_{T}\rangle$& $\langle
F_{L}\rangle$ \\
 \hline
   $B^+\rightarrow\bar D^0 e^+\nu_e$ & $2.53\times10^{-2}$ & $-0.98\times10^{-6}$ & $-1.50$ & $1.00$ & $-1.01\times10^{-3}$ & \quad\\
$B^+\to \bar D^0\mu^+\nu_\mu$&$2.52\times10^{-2}$ &$-0.013$ & $-1.46$ & $0.96$ &$-0.19$ &
    \\
    $B^+\to\bar D^0\tau^+\nu_\tau$ &$0.68\times10^{-2}$&$-0.37$ &
                                                                  $-0.30$ & $-0.24$ &$-0.85$ &\\
     $B^+\rightarrow\bar D^{*0} e^+\nu_e$& $6.81\times10^{-2}$ & $-0.22$ & $-0.48$ & $1.00$ & $-0.34\times10^{-3}$ & 0.55\\
 $B^+\to\bar D^{*0}\mu^+\nu_\mu$  &$6.77\times10^{-2}$  &$-0.23$ & $-0.47$ & $0.98$ &$-0.061$&0.55
    \\
  $B^+\to\bar D^{*0}\tau^+\nu_\tau$ &$1.56\times10^{-2}$  &$-0.32$ & $-0.060$ & $0.48$ &$-0.12$ &0.47
    \\
 $B^+\rightarrow\pi^0 e^+\nu_e$&$7.20\times10^{-5}$&$-0.28\times10^{-6}$&$-1.50$&$1.00$&$-0.46\times10^{-3}$&\\

   $B^+\to \pi^0\mu^+\nu_\mu$ &$0.72\times10^{-4}$&$-0.004$ & $-1.49$ & $0.99$ &$-0.09$ &
    \\
    $B^+\to \pi^0\tau^+\nu_\tau$&$0.45\times10^{-4}$ &$-0.22$ &
                                                                $-0.82$ & $0.42$ &$-0.72$ &\\
     $B^+\rightarrow\rho e^+\nu_e$ & $1.74\times10^{-4}$&$-0.50$&$0.042$&$1.00$&$-0.07\times10^{-3}$&0.31\\
   $B^+\to \rho^0\mu^+\nu_\mu$ &$1.73\times10^{-4}$&$-0.51$ & $0.047$ & $0.99$ &$-0.011$&0.31
    \\
    $B^+\to \rho^0\tau^+\nu_\tau$ &$0.97\times10^{-4}$&$-0.54$ & $0.14$ & $0.60$ &$0.095$ &0.31
    \\
      $B^+\rightarrow\eta e^+\nu_e$&$4.24\times10^{-5}$&$-0.37\times10^{-6}$&$-1.50$&$1.00$&$-0.60\times10^{-3}$&\\
  $B^+\to \eta\mu^+\nu_\mu$ &$0.42\times10^{-4}$&$-0.006$ & $-1.48$ & $0.98$ &$-0.12$ &
    \\
    $B^+\to \eta\tau^+\nu_\tau$&$0.26\times10^{-4}$ &$-0.27$ & $-0.67$ &
                                                                     $0.21$ &$-0.83$ &\\
 $B^+\to \eta'e^+\nu_e$&$3.17\times10^{-5}$&$-0.43\times10^{-6}$&$-1.50$&$1.00$&$-0.66\times10^{-3}$&\\
    $B^+\to \eta'\mu^+\nu_\mu$ &$0.31\times10^{-4}$&$-0.007$ & $-1.48$ & $0.98$ &$-0.13$ &
    \\
     $B^+\to \eta'\tau^+\nu_\tau$&$0.17\times10^{-4}$ &$-0.30$ &
                                                                 $-0.59$ & $0.14$ &$-0.84$ &\\
     $B^+\rightarrow\omega e^+\nu_e$&$1.71\times10^{-4}$&$-0.43$&$-0.17$&$1.00$&$-0.12\times10^{-3}$&0.41\\
   $B^+\to \omega\mu^+\nu_\mu$ &$1.71\times10^{-4}$&$-0.43$ & $-0.16$ & $0.99$ &$-0.02$&0.41
    \\
    $B^+\to \omega\tau^+\nu_\tau$ &$0.97\times10^{-4}$&$-0.49$ & $0.009$ & $0.59$ &$-0.007$ &0.40
    \\
    \end{tabular}
  \end{ruledtabular}
\end{table}

  \begin{table}
\caption{The branching ratios, FB asymmetry and polarization parameters
  of semileptonic $B_s$ decays in RQM.}
\label{asBs}
\begin{ruledtabular}
  \begin{tabular}{ccccccc}
   Decay&Br&$\langle A_{FB}\rangle$  & $\langle C^\ell_{F}\rangle$&
$\langle P^\ell_{L}\rangle$  &$\langle P^\ell_{T}\rangle$& $\langle
F_{L}\rangle$ \\
 \hline
  $B_s\rightarrow D_s^- e^+\nu_e$&$2.12\times10^{-2}$&$-0.97\times10^{-6}$&$-1.50$&1.00&$-1.02\times10^{-3}$&\\
$B_s\to D_s^-\mu^+\nu_\mu$ &$2.12\times10^{-2}$&$-0.013$ & $-1.46$ & $0.96$ &$-0.19$ &
    \\
    $B_s\to D_s^-\tau^+\nu_\tau$ &$0.61\times10^{-2}$&$-0.36$ &
                                                                  $-0.30$ & $-0.27$ &$-0.85$ &\\
    $B_s\rightarrow D_s^{*-} e^+\nu_e$&$5.06\times10^{-2}$&$-0.26$&$-0.35$&$1.00$&$-0.23\times10^{-3}$&$0.49$\\
 $B_s\to D_s^{*-}\mu^+\nu_\mu$ &$5.05\times10^{-2}$   &$-0.27$ & $-0.33$ & $0.99$ &$-0.040$&0.49
    \\
  $B_s\to D_s^{*-}\tau^+\nu_\tau$  &$1.23\times10^{-2}$ &$-0.32$ & $-0.040$ & $0.53$ &$-0.035$ &0.42
    \\
    $B_s\rightarrow K^- e^+\nu_e$&$15.6\times10^{-5}$&$-0.39\times10^{-6}$&$-1.50$&$1.00$&$-0.56\times10^{-3}$&\\
    $B_s\to K^-\mu^+\nu_\mu$&$1.55\times10^{-4}$ &$-0.006$ & $-1.48$ & $0.98$ &$-0.11$ &
    \\
    $B_s\to K^-\tau^+\nu_\tau$ &$0.91\times10^{-4}$&$-0.24$ &
                                                                $-0.77$ & $0.35$ &$-0.75$ &\\
    $B_s\rightarrow K^{*-} e^+\nu_e$&$3.29\times10^{-4}$&$-0.37$&$-0.23$&$1.00$&$-0.14\times10^{-3}$&0.44\\
 $B_s\to K^{*-}\mu^+\nu_\mu$ &$3.29\times10^{-4}$   &$-0.38$ & $-0.22$ & $0.99$ &$-0.025$&0.44
    \\
  $B_s\to K^{*-}\tau^+\nu_\tau$ &$1.82\times10^{-4}$  &$-0.44$ & $-0.032$ & $0.63$ &$-0.025$ &0.42
  \\
  \end{tabular}
\end{ruledtabular}
\end{table}

\begin{table}
\caption{The branching ratios, FB asymmetry and polarization parameters
  of semileptonic $B_c$ decays in RQM.}
\label{asBc}
\begin{ruledtabular}
  \begin{tabular}{ccccccc}
    Decay&Br&$\langle A_{FB}\rangle$  & $\langle C^\ell_{F}\rangle$&
$\langle P^\ell_{L}\rangle$  &$\langle P^\ell_{T}\rangle$& $\langle
F_{L}\rangle$ \\
 \hline
   $B_c^+\to \eta_ce^+\nu_e$ &$0.42\times10^{-2}$&$-8.6\times10^{-7}$ & $-1.5$ & $1.0$ &$-1.03\times10^{-3}$ &
    \\
   $B_c^+\to \eta_c\mu^+\nu_\mu$ &$0.42\times10^{-2}$&$-0.012$ & $-1.46$ & $0.96$ &$-0.20$ &
    \\
    $B_c^+\to \eta_c\tau^+\nu_\tau$ &$0.16\times10^{-2}$&$-0.35$ &
                                                                   $-0.24$ & $-0.39$ &$-0.82$ &\\
  $B_c^+\to J/\psi e^+\nu_e$ &$1.31\times10^{-2}$   &$-0.19$ & $-0.23$ & $1.0$ &$-0.15\times10^{-3}$&0.44
    \\
 $B_c^+\to J/\psi\mu^+\nu_\mu$ &$1.30\times10^{-2}$   &$-0.19$ & $-0.23$ & $0.99$ &$-0.028$&0.44
    \\
  $B_c^+\to J/\psi\tau^+\nu_\tau$  &$0.37\times10^{-2}$ &$-0.23$ & $-0.032$ & $0.56$ &$-0.069$ &0.40
  \\
    $B_c^+\to\bar D^0e^+\nu_e$&$0.33\times10^{-4}$ &$-1.0\times10^{-7}$ & $-1.5$ & $1.0$ &$-0.43\times10^{-3}$&
    \\
   $B_c^+\to\bar D^0\mu^+\nu_\mu$&$0.33\times10^{-4}$ &$-0.002$ & $-1.49$ & $0.99$ &$-0.088$ &
    \\
    $B_c^+\to\bar D^0\tau^+\nu_\tau$ &$0.28\times10^{-4}$&$-0.257$ &
                                                                     $-0.72$ & $0.25$ &$-0.83$ &\\
  $B_c^+\to\bar D^{*0}e^+\nu_e$ &$0.84\times10^{-4}$   &$-0.45$ & $0.10$ & $1.0$ &$6.7\times10^{-5}$&0.29
    \\
 $B_c^+\to\bar D^{*0}\mu^+\nu_\mu$ &$0.84\times10^{-4}$   &$-0.45$ & $0.10$ & $1.0$ &$0.014$&0.29
    \\
  $B_c^+\to\bar D^{*0}\tau^+\nu_\tau$ &$0.55\times10^{-4}$  &$-0.42$ & $0.072$ & $0.78$ &$0.22$ &0.29
    \\
 $B_c^+\to B_se^+\nu_e$ &$0.92\times10^{-2}$&$-1.1\times10^{-5}$ & $-1.5$ & $1$ &$-0.004$ &
    \\
    $B_c^+\to B_s\mu^+\nu_\mu$ &$0.90\times10^{-2}$&$-0.093$ & $-1.21$ & $0.73$ &$-0.57$ &\\
 $B_c^+\to B_s^*e^+\nu_e$ &$2.30\times10^{-2}$   &$-0.15$ & $-0.29$ & $1$ &$-7.3\times10^{-4}$&0.46
    \\
  $B_c^+\to B_s^*\mu^+\nu_\mu$  &$2.20\times10^{-2}$ &$-0.17$ &
                                                                  $-0.23$ & $0.93$ &$-0.09$ &0.46
    \\
    $B_c^+\to B e^+\nu_e$ &$4.94\times10^{-4}$&$-9.0\times10^{-6}$ & $-1.5$ & $1$ &$-0.004$ &
    \\
    $B_c^+\to B\mu^+\nu_\mu$ &$4.79\times10^{-4}$&$-0.080$ & $-1.25$ & $0.76$ &$-0.53$ &\\
 $B_c^+\to B^*e^+\nu_e$ &$17.9\times10^{-4}$   &$-0.17$ & $-0.27$ & $1$ &$-6.2\times10^{-4}$&0.45
    \\
  $B_c^+\to B^*\mu^+\nu_\mu$  &$17.2\times10^{-4}$ &$-0.18$ & $-0.22$ & $0.94$ &$-0.09$ &0.45 \\
\end{tabular}
\end{ruledtabular}
\end{table}

\section{Comparison of RQM and CLFQM predictions}

In this section we compare predictions for semileptonic decay form
factors, decay rates and FB asymmetries in RQM and the other popular quark
model CLFQM \cite{v} and confront them with available experimental data and
lattice calculations. The form factors of semileptonic $B$ meson decays in  CLFQM are given in Refs.~\cite{v,zkgdlw}.
In the following tables, the results in CLFQM are mainly from Ref.~\cite{zkgdlw}.

In Table~\ref{ff1c} we present predictions for
the form factors  of the weak $B$ and $B_s$  meson transitions at
$q^2=0$ and   $q^2=q^2_{\rm max}$ in RQM and CLFQM. We find that the
form factor values at $q^2=0$ are close in both models for all
considered decays. The values of form factors at  $q^2=q^2_{\rm max}$
are also close for $B$ and $B_s$ decays to heavy $D^{(*)}$ and
$D_s^{(*)}$ mesons. They are consistent with the recent  lattice QCD calculations
\cite{flm,mdkl,hd,mnsv,hpqcdbc}. On the other hand, the values of form factors for
heavy-to-light transitions at  $q^2=q^2_{\rm max}$ significantly
differ in the considered models. The values of form factors at
this kinematic point in RQM are substantially higher than the ones in CLFQM.

\label{sec:comp}
\begin{table}
\caption{Comparison of form factors of the weak $B$ and $B_s$   meson
  transitions in RQM and CLFQM. }
\label{ff1c}
\begin{ruledtabular}
\begin{tabular}{cccccc}
Transition& Form factor &  \multicolumn{2}{c}{$F(0)$}&
                                                  \multicolumn{2}{c}{$F(q^2_{\rm
                                                  max})$ }\\
   \cline{3-4} \cline{5-6}
    &&RQM&CLFQM&RQM&CLFQM\\
\hline
  $B\to D$         &$f_+$ &$0.696$ &0.67& $1.24$ & 1.21 \\
   & $f_0$  &$0.696$ &0.67& $0.82$ & 0.91 \\ \hline
$B\to D^*$   &$V$&$0.915$&0.77& $1.38$& $1.34$\\
     &$A_0$&$0.814$&0.68&$1.21$&$1.15$\\
     &$A_1$&$0.730$&0.65& $0.83$& $0.84$\\
     &$A_2$&$0.627$&0.62& $1.02$& $0.99$ \\ \hline
  $B\to \pi$   &$f_+$ &$0.217$ &0.25& $10.9$ & 1.27\\
     & $f_0$  &$0.217$ &0.25& $1.32$ & $0.82$ \\ \hline
$B\to \rho$   &$V$&$0.295$&0.29& $2.80$& $1.06$ \\
     &$A_0$&$0.231$&0.32&$2.19$&$1.00$\\
     &$A_1$&$0.269$&0.24& $0.439$& $0.53$ \\
     &$A_2$&$0.282$& 0.22&$1.92$&$0.70$\\ \hline
$B\to \eta$   &$f_+$ &$0.194$ &0.22& $2.75$ &1.31 \\
     & $f_0$  &$0.194$ &0.22& $0.52$ & $0.52$ \\ \hline
$B\to \eta'$   &$f_+$ &$0.187$ &0.18& $1.34$ & $0.75$ \\
   & $f_0$  &$0.187$ &0.18& $0.32$ &  $0.35$ \\ \hline
$B\to \omega$   &$V$&$0.263$& 0.27&$2.47$& $0.88$ \\
     &$A_0$&$0.244$&0.28&$1.96$&$0.60$\\
     &$A_1$&$0.232$&0.23 &$0.46$& $0.53$ \\
     &$A_2$&$0.229$&0.21& $0.96$& $0.63$ \\ \hline
  $B_s\to D_s$         &$f_+$ &$0.663$ &0.67& $1.23$ & 1.20 \\
   & $f_0$  &$0.663$ &0.67& $0.87$ &  $0.92$ \\ \hline
$B_s\to D_s^*$   &$V$&$0.925$&0.75& $1.48$ &$1.30$\\
     &$A_0$&$0.625$&0.66&$1.03$&$1.13$\\
     &$A_1$&$0.668$&0.62& $0.82$& $0.84$ \\
     &$A_2$&$0.723$& 0.57&$1.09$& $0.94$ \\ \hline
   $B_s\to K$         &$f_+$ &$0.284$ &0.23& $5.42$ & $0.42$ \\
   & $f_0$  &$0.284$ &0.23& $0.459$ & $0.64$ \\ \hline
$B_s\to K^*$   &$V$&$0.291$&0.23& $3.06$& $0.32$ \\
     &$A_0$&$0.289$&0.25&$2.10$&$0.34$\\
     &$A_1$&$0.287$&0.19 &$0.581$& $0.36$ \\
     &$A_2$&$0.286$&0.16 &$0.953$& $0.25$ \\
\end{tabular}
\end{ruledtabular}
\end{table}

\begin{figure}
\centering
  \includegraphics[width=13cm]{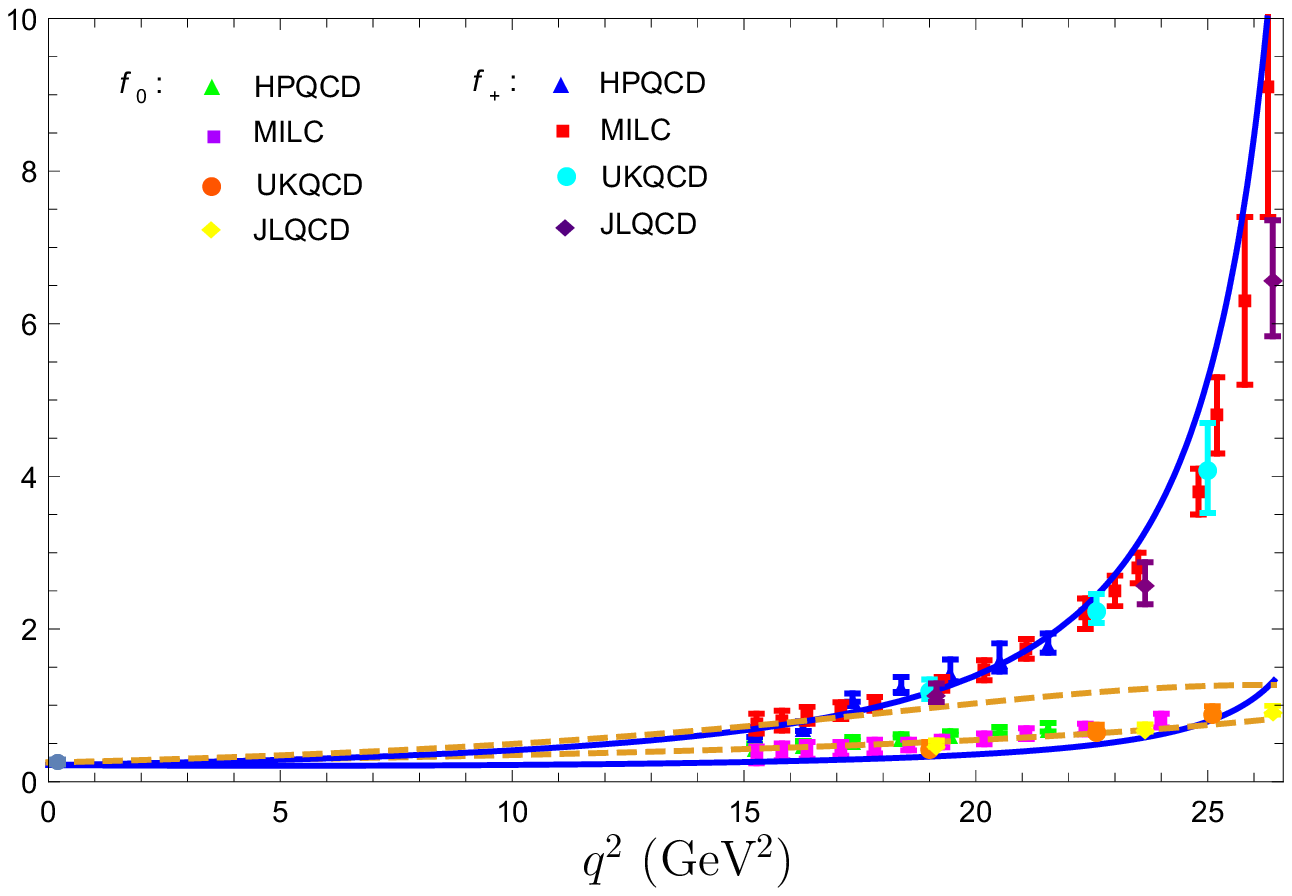}
\caption{Comparison of theoretical predictions for the form factors of
  the weak $B\to \pi$ transitions. Model results: the RQM form factors
  are given by solid blue lines, and the CLFQM form factors are plotted by orange dashed
  lines. Upper curves correspond to $f_+(q^2)$ and the lower ones to
  $f_0(q^2)$. Lattice results are plotted: HPQCD \cite{hpqcd}  by triangles (blue $f_+(q^2)$, green  $f_0(q^2)$)  with error bars, FNAL/MILC \cite{fnal,fnal1} by
  squares (red $f_+(q^2)$, magenta $f_0(q^2)$) with error bars, RBC/UKQCD \cite{ukqcd}
  by filled circles (cyan $f_+(q^2)$, orange $f_0(q^2)$)   with error
  bars, and JLQCD \cite{jlqcd} by diamonds (purple  $f_+(q^2)$, yellow $f_0(q^2)$) .   }
\label{ff}
\end{figure}

\begin{figure}
\centering
  \includegraphics[width=13cm]{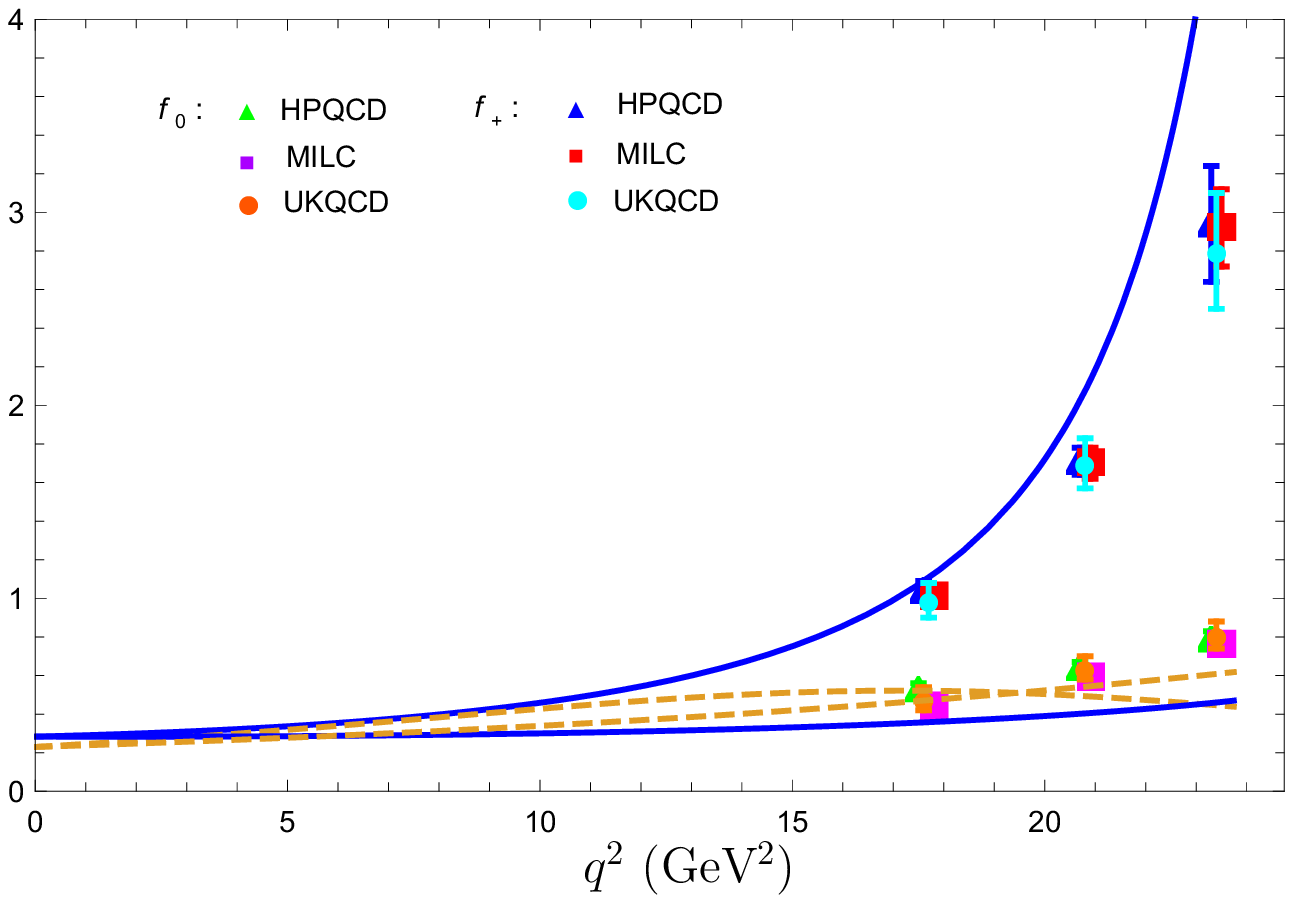}
\caption{Same as in Fig.~\ref{ff}, but for the form factors of
  the weak $B_s\to K$ transitions. For the orange dashed lines, the upper one below $q^2<15$ GeV$^2$ corresponds to $f_+(q^2)$, and the lower one $f_0(q^2)$. HPQCD, MILC and UKQCD data are from Refs.~\cite{hpqcd}, \cite{milc} and \cite{ukqcd}, respectively. }
\label{ffBs}
\end{figure}

 In Fig.~\ref{ff} we plot the form factors $f_+(q^2)$ and
$f_0(q^2)$ of the $B\to \pi$ semileptonic transitions in the whole kinematic
range. Predictions of RQM are plotted by solid blue lines and  those
of CLFQM are given by dashed orange lines.  From this plot we see that
the form factor  $f_+(q^2)$ in RQM rapidly grows for
$q^2>15$ GeV$^2$ up to the value
$f_+(q^2_{\rm max})=10.9$ , while this form factor in CLFQM reaches
only  the value $f_+(q^2_{\rm max})=1.27$ which is almost an order of
magnitude lower. In the same figure we also give  lattice QCD
results \cite{hpqcd,fnal,fnal1,ukqcd,jlqcd,msv} which are available in the high
$q^2>15$ GeV$^2$ region. The form factors of the $B_s\to K$ semileptonic
transitions in RQM and CLFQM in comparison with lattice QCD data
\cite{hpqcd1,milc,ukqcd,msv} are plotted  in Fig.~\ref{ffBs}. The model
predictions are again significantly different for the form factor
$f_+(q^2)$. It rapidly grows in RQM, contrary in  CLFQM it slowly increases
reaching the maximum value at $q^2$ of about 15~GeV$^2$ and then slowly
decreases. Note that in CLFQM the value of $f_+(q^2_{\rm max})$ is
even smaller than the value of $f_0(q^2_{\rm max})$.
From these plots we see that predictions of RQM for the form
factor $f_+$ are in much better agreement with lattice data, which
also favor its rapid growth with $q^2$.

The main reason of the discrepancy between the $q^2$ dependence of the
heavy-to-light form factors  in Figs.~\ref{ff} and \ref{ffBs}
 in  RQM and   CLFQM is the following. In RQM the form
factors are directly calculated in the whole $0\le q^2\le q^2_{\rm
  max}$ region without any extrapolations. As the result the pole
structure of the form factors $f_+(q^2)$, $V(q^2)$ and $A_0(q^2)$ in
Eq.~(\ref{fitfv}) is explicitly obtained already in the impulse
approximation. The inclusion of the negative-energy contributions
($Z$-diagrams) enhances the growth of these form factors for the
heavy-to-light decays in the high $q^2$ region, bringing results in a
better agreement with lattice data.   Contrary, in CLFQM the $q^+=0$
frame (light-cone gauge) is used, resulting in the determination of
the form factors for $q^2<0$, and then they are extrapolated to the
$q^2>0$ region. As has been shown in 
Refs.~\cite{j,lb}, for the $q^+=0$ case, the nonvalence contributions ($Z$-
diagrams) are highly suppressed, leading to a relativistic picture of a
meson as only the valence $q\bar q$ structure. Such extrapolation
works well for the heavy-to-heavy transitions with a rather small
$q^2$ range, but fails to reproduce the pole
structure of the form factors $f_+(q^2)$, $V(q^2)$ and $A_0(q^2)$ for
the heavy-to-light transitions which have a very broad $q^2$ range. 
One may indeed work in the $q^+ \neq 0$ frame, as a more realistic
case for the time-like region, then $Z$- diagrams will be inevitable,
and moreover for a larger $q^2$ region the contribution of $Z$- diagrams
becomes more important. This should be the main origin for the
inconsistency between RQM and CLFQM for the heavy-to-light decays, for
which we used the $B\to \pi$ and $B_s\to K$ transitions as examples.

\begin{figure}
\centering
  \includegraphics[width=13cm]{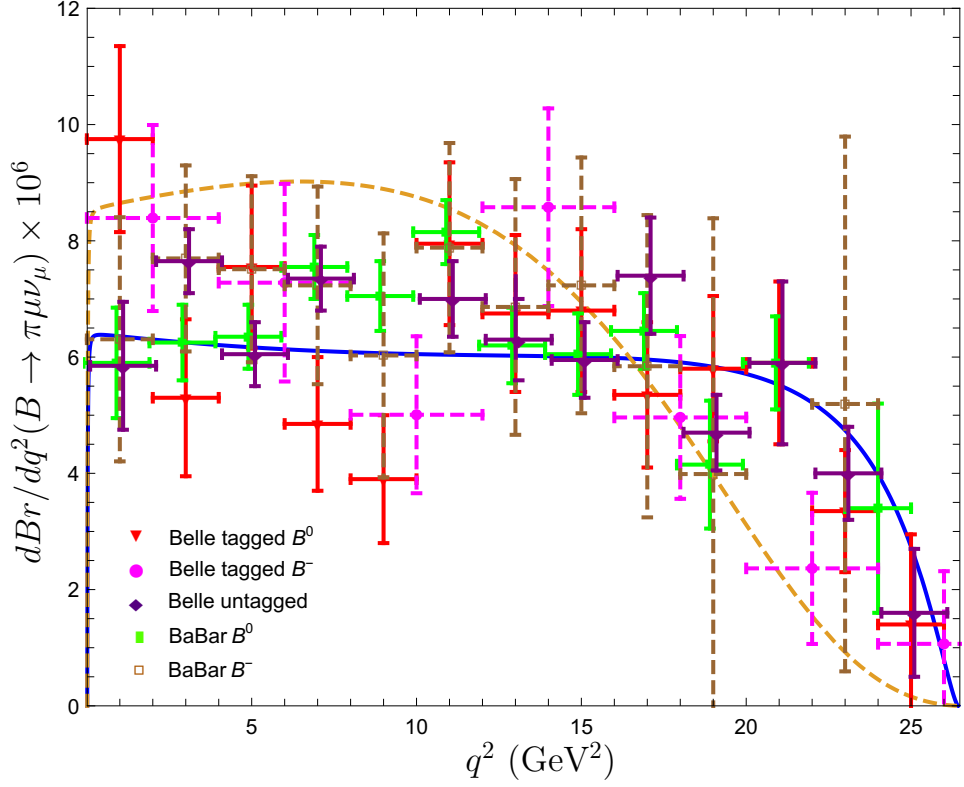}
\caption{Differential branching fractions of the semileptonic
  $B\to \pi\mu\nu_\mu$ decay.  Comparison of theoretical predictions (RQM -- solid blue line,  CLFQM -- orange dashed
  line) with available experimental data \cite{belle1,belle2,babar}. }
\label{brpi}
\end{figure}

\begin{figure}
\centering
  \includegraphics[width=13cm]{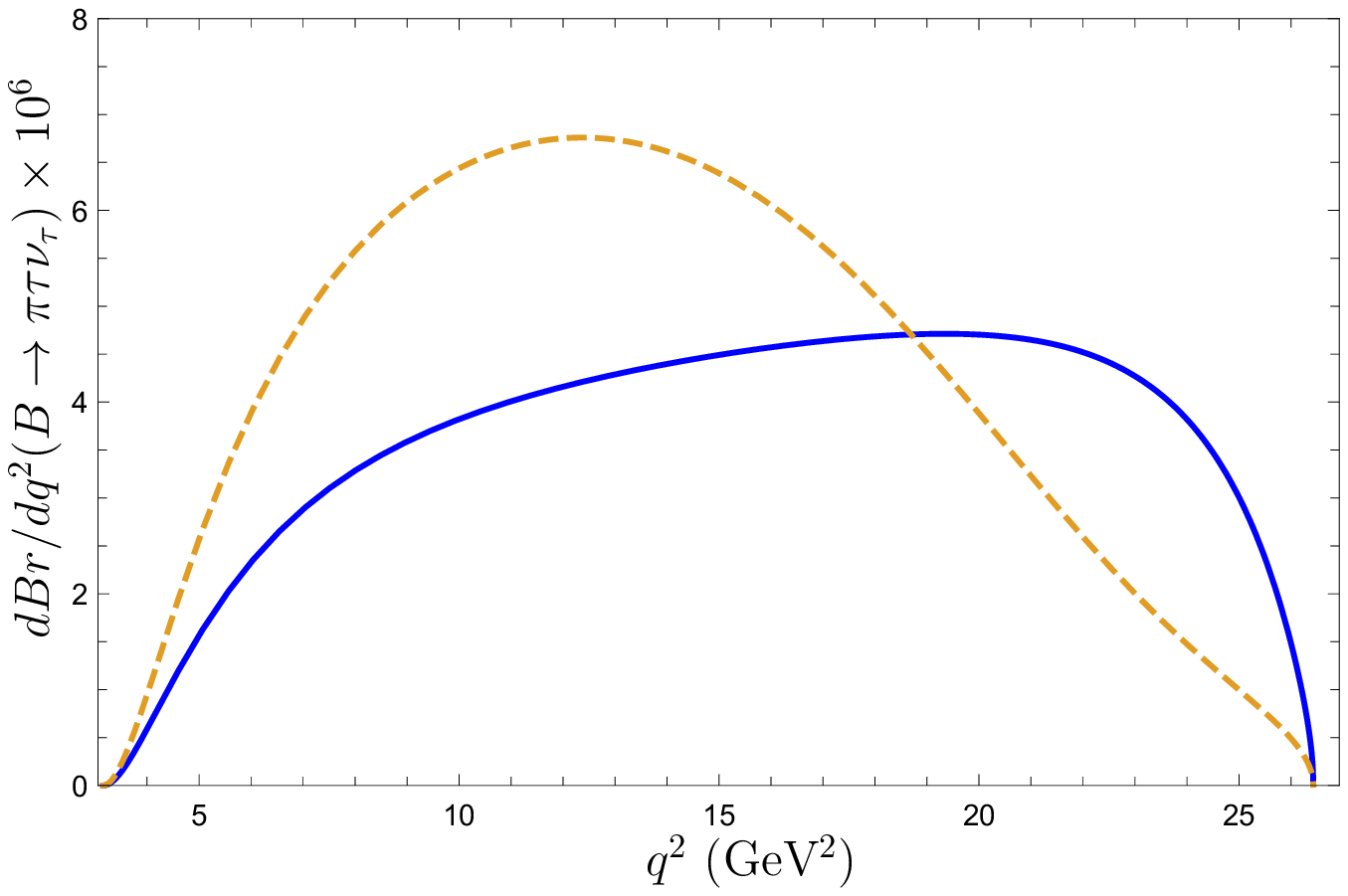}
\caption{Differential branching fractions of the semileptonic
  $B\to \pi\tau\nu_\tau$ decay.  Comparison of theoretical predictions (RQM -- solid blue lines,  CLFQM -- orange dashed
  lines). }
\label{brpitau}
\end{figure}

In Table~\ref{compexp} we present the comparison of the RQM and CLFQM model
predictions for the semileptonic $B$ and $B_s$ decay branching ratios
with available experimental data \cite{pdg}. We see   that the
branching ratios, which are integrated quantities, are   consistent
with each other and experimental data. However, the   differential
distributions can differ substantially due to the difference in the
$q^2$ dependence of the form factors. As an example, in
Figs.~\ref{brpi} and \ref{brpitau} we plot the differential branching
fractions in both models for $B\to\pi\mu\nu_\mu$ and
$B\to\pi\tau\nu_\tau$ decays, respectively. From these plots we see
that differential distributions, which are plotted by solid blue line
for RQM and orange dashed lines for CLFQM,  differ significantly. In
Fig.~\ref{brpi} we also show the available experimental data from
Belle \cite{belle1,belle2} and BaBar \cite{babar} Collaborations. It is not possible to  unambiguously
distinguish between models since the data have large uncertainties,
but the shape of the differential distribution predicted by RQM is in
better agreement with the averaged data.

\begin{table}
\caption{Comparison of theoretical predictions for the semileptonic decay
  branching ratios with available experimental data \cite{pdg}.}
\label{compexp}
\begin{ruledtabular}
  \begin{tabular}{lccc}
    Decay& \multicolumn{2}{c}{Theory}& Experiment \\
    \cline{2-3}
    &RQM&CLFQM&\\
    \hline
$B^+\to\bar D^0\ell^+\nu_\ell$ &2.52\%&2.22\%&$(2.35\pm0.09)$\%\\
 $B^+\to\bar D^0\tau^+\nu_\tau$ &$6.8\times 10^{-3}$&$6.7\times 10^{-3}$&$(7.7\pm
  2.5)\times 10^{-3}$\\
 $B^+\to\bar D^{*0}\ell^+\nu_\ell$ &6.77\%&5.70\%&$(5.66\pm0.22)$\%\\
    $B^+\to\bar D^{*0}\tau^+\nu_\tau$ &1.56\%&1.40\%&$(1.88\pm0.20)$\%\\
  $B^+\to\pi^0\ell^+\nu_\ell$ &$7.2\times 10^{-5}$&$7.8\times 10^{-5}$&$(7.80\pm
                                                         0.27)\times 10^{-5}$\\
   $B^+\to \eta\ell^+\nu_\ell$ &$4.2\times 10^{-5}$&$5.4\times 10^{-5}$&$(3.9\pm
                                                         0.5)\times 10^{-5}$\\
    $B^+\to\eta'\ell^+\nu_\ell$ &$3.1\times 10^{-5}$&$2.6\times 10^{-5}$&$(2.3\pm
                                                      0.8)\times 10^{-5}$\\
    $B^+\to \rho^0\ell^+\nu_\ell$ &$1.73\times 10^{-4}$&$2.19\times 10^{-4}$&$(1.58\pm
                                                         0.11)\times 10^{-4}$\\
    $B^+\to\omega\ell^+\nu_\ell$ &$1.71\times 10^{-4}$&$2.07\times 10^{-4}$&$(1.19\pm
                                                        0.09)\times 10^{-5}$\\
    $B^0\to D^-\ell^+\nu_\ell$ &2.33\%&2.05\%&$(2.31\pm0.10)$\%\\
 $B^0\to D^-\tau^+\nu_\tau$ &$0.63\%$&$0.62\%$&$(1.08\pm
  0.23)\%$\\
 $B^0\to D^{*-}\ell^+\nu_\ell$ &6.28\%&5.28\%&$(5.06\pm0.12)$\%\\
    $B^0\to D^{*-}\tau^+\nu_\tau$ &1.45\%&1.30\%&$(1.57\pm0.09)$\%\\
    $B^0\to\pi^-\ell^+\nu_\ell$ &$1.33\times 10^{-4}$&$1.44\times 10^{-4}$&$(1.50\pm
                                                         0.06)\times 10^{-4}$\\
     $B^0\to\pi^-\tau^+\nu_\tau$ &$0.84\times 10^{-4}$&$0.98\times 10^{-4}$&$<2.5\times
                                                        10^{-4}$\\
 $B^0\to\rho^-\ell^+\nu_\ell$ &$3.2\times 10^{-4}$&$4.1\times 10^{-4}$&$(2.94\pm
                                                         0.21)\times 10^{-4}$\\
 $B_s\to D_s^-\mu^+\nu_\mu$ &2.12\%&2.05\%&$(2.52\pm0.24)$\%\\
    $B_s\to D_s^{*-}\mu^+\nu_\mu$ &5.05\%&5.05\%&$(5.4\pm0.5)$\%\\
     $B_s\to K^-\mu^+\nu_\mu$&$1.55\times10^{-4}$ &$1.01\times10^{-4}$&$(1.06\pm0.10)\times10^{-4}$\\
  \end{tabular}
\end{ruledtabular}
\end{table}

In Table~\ref{ggcomp} we  give the ratios of the decay rates with
$\tau$ and $\mu$ leptons, ${\cal R}(F)=\Gamma(B\to
F\tau\nu_\tau)/\Gamma(B\to F\mu\nu_\mu)$, predicted by RQM, CLFQM and
lattice in comparison with available experimental data. Such
comparison provides the test of the lepton universality and deviations
of the standard model predictions from experimental data can
indicate the possible contributions of new physics. We see that
theoretical predictions are consistent with each other and are lower than
experimental data by about $1\sim 3\sigma$.

\begin{table}
\caption{Ratios of the decay rates with $\tau$ and $\mu$ leptons
  ${\cal R}(F)=\Gamma(B\to F\tau\nu_\tau)/\Gamma(B\to F\mu\nu_\mu)$ in comparison with available lattice or 
  experimental data, cf. Ref.~\cite{bsrw} and references therein. }
\label{ggcomp}
\begin{ruledtabular}
\begin{tabular}{ccccccc}
Transition& \multicolumn{3}{c}{Theory} & \multicolumn{3}{c}{Experiment}
                \\  \cline{2-4}\cline{5-7}
   & RQM&CLFQM&Lattice/SM analysis \cite{bsrw}&PDG \cite{pdg}&HFLAV
                                                   \cite{hflav}&\cite{bsrw}\\
                                                     \hline
  $B\to D$ &$0.271$&0.302&0.298(3) & $0.429(82)(52)$($B^+$) &0.339(26)(14)&0.337(30)\\
                   &   &&&$0.469(84)(53)$($B^0$) \\
  $B\to D^{*}$  &$0.231$&0.246&0.250(3) &  $0.335(34)$($B^+$)&0.295(10)(10)&0.298(14)\\
                    &      & &&$0.309(16)$($B^0$)\\ \hline
$B\to \pi$ &$0.631$&0.680&0.641(16) &&&1.05(51)\\
  $B\to \rho$  &$0.561$&0.543&0.535(8) \\ \hline
  $B\to \eta$ &$0.629$&0.611 \\
  $B\to \eta'$ &$0.544$&0.538 \\
 $B\to \omega$   &$0.566$&0.531&0.546(15) \\ \hline
$B_s\to D_s$ &$0.287$&0.298&0.297(3)  \\
 $B_s\to D_s^*$  &$0.244$&0.248 &0.247(8) \\ \hline
  $B_s\to K$ &$0.588$&0.673 \\
    $B_s\to K^*$  &$0.553$&0.520 \\ \hline
  $B_c\to \eta_c$ &$0.373$  \\
  $B_c\to J/\psi$&$0.284$&&0.2582(38)& $0.71(17)(18)$ \\ \hline
  $B_c\to D$ &$0.833$ \\
  $B_c\to D^*$&$0.656$ \\
\end{tabular}
\end{ruledtabular}
\end{table}

For the semileptonic $\bar B\to D^*\tau^-\bar\nu_\tau$  decay the $\tau$-lepton
 polarization and longitudinal polarization fraction of the final
 vector $D^*$ meson were recently measured experimentally
 \cite{bellepol,Belle:2016dyj}. We compare  RQM,  CLFQM and  lattice predictions
 with these data in Table~\ref{taucomp}. The theoretical predictions
 for the $\tau$-lepton longitudinal polarization $\langle
 P^\tau_L\rangle$ agree with each other and experimental data within
 large error bars. The experimental value for the longitudinal
 polarization fraction of the final vector $D^*$ meson  $\langle
 F_L\rangle$ is about $2\sigma$ higher than theoretical predictions.

\begin{table}
\caption{Comparison of theoretical predictions for the $\tau$-lepton
  polarization and longitudinal polarization fraction of the final
  vector meson with experimental data \cite{bellepol,Belle:2016dyj}. }
\label{taucomp}
\begin{ruledtabular}
\begin{tabular}{ccccccccc}
Decay& \multicolumn{4}{c}{$\langle P^\tau_L\rangle$} &
                                                       \multicolumn{4}{c}{$\langle
                                                       F_L\rangle$}
                \\  \cline{2-5}\cline{6-9}
   & RQM&CLFQM&Lattice \cite{msv}&Experiment &RQM&CLFQM&Lattice \cite{msv}&Experiment\\
  \hline
  $\bar B\to D^*\tau^-\bar\nu_\tau$ &$-0.48$&$-0.51$&$-0.529(7)$ & $-0.38(51)(^{21}_{10})$ &0.47&0.45&0.414(12)&0.60(8)(4)\\
  $\bar B_s\to D_s^{*}\tau^-\bar\nu_\tau$  &$-0.53$&$-0.51$&$-0.520(12)$ &  &0.42&0.45&0.404(16)&
\end{tabular}
\end{ruledtabular}
\end{table}

In Table~\ref{pikcomp} we compare  RQM and CLFQM predictions with lattice data for
 the FB asymmetry and lepton polarization for the
 heavy-to-light semileptonic $B$ and $B_s$ decays to
 light pseudoscalar mesons. Such comparison is important since RQM and
 CLFQM have significantly different dependence  of the form
 factor $f_+(q^2)$ for large values of $q^2$. From this table we see
 that the lepton polarization $\langle P^\tau_L\rangle$ is very
 sensitive to the employed model. Its values are  substantially
 smaller in CLFQM, and for  $ B_s\to K\tau^+\nu_\tau$ they have
 even opposite signs. In general RQM results for these observables
 agree better with the combined lattice values.

\begin{figure}
\centering
  \includegraphics[width=8cm]{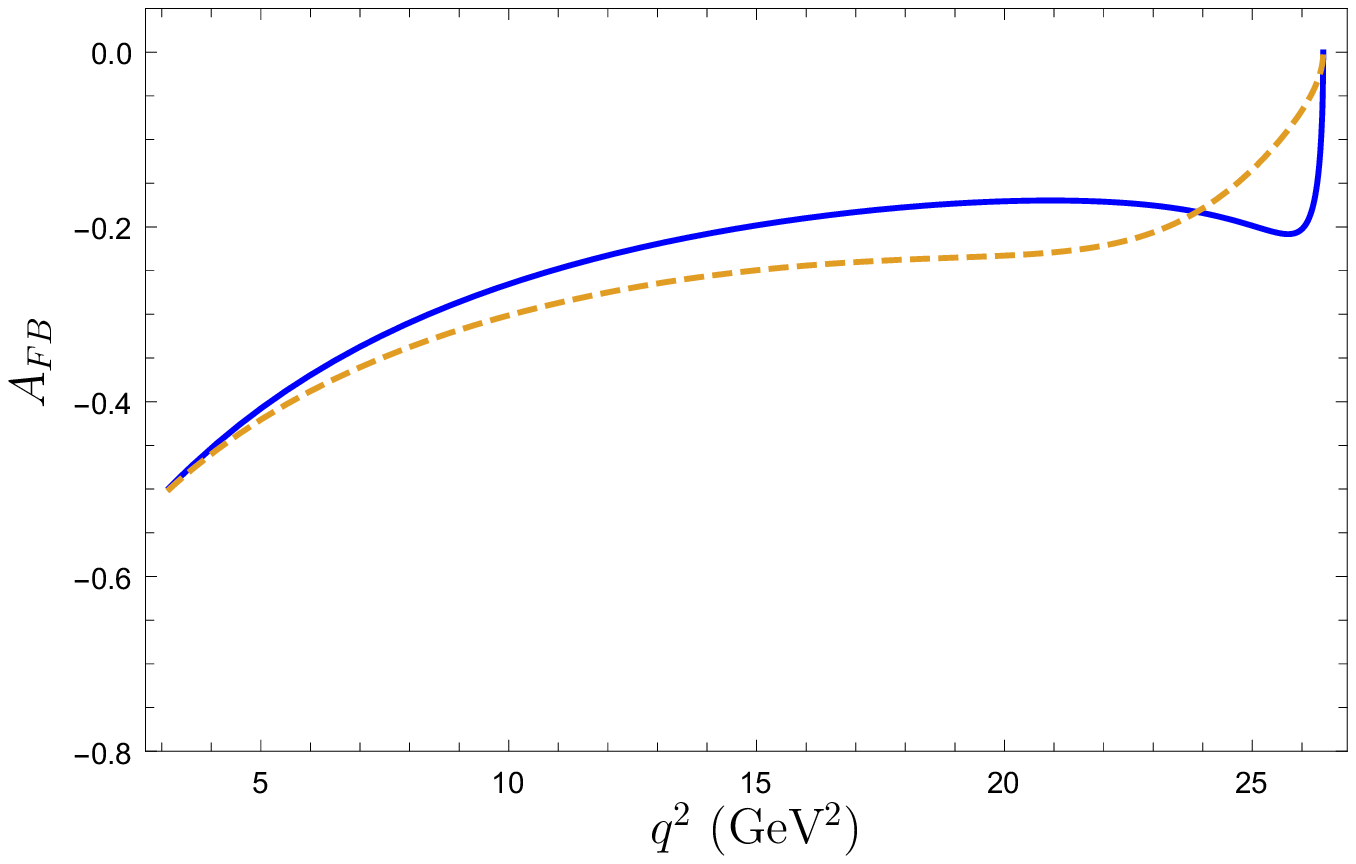}\ \
  \ \includegraphics[width=8cm]{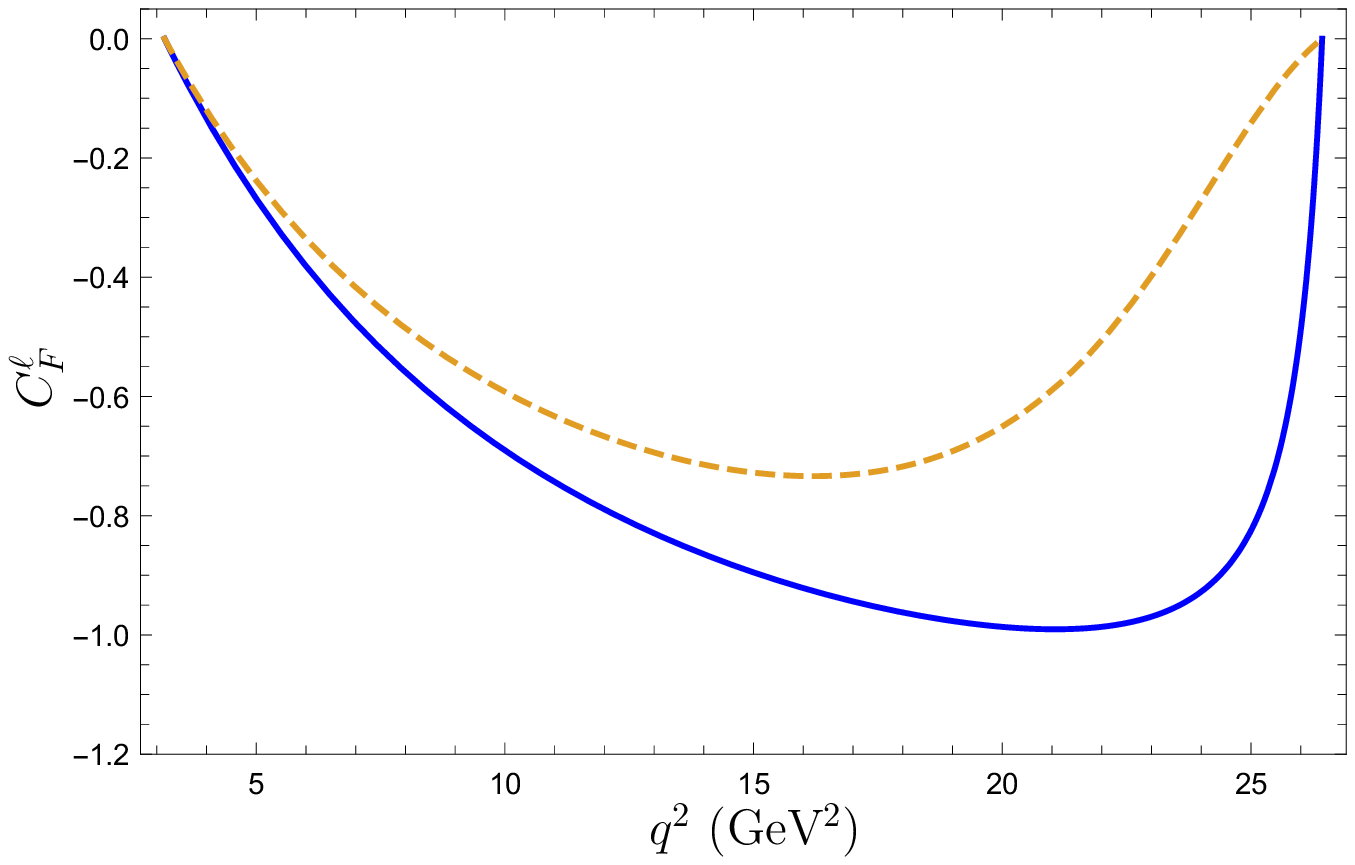}\\
  \includegraphics[width=8cm]{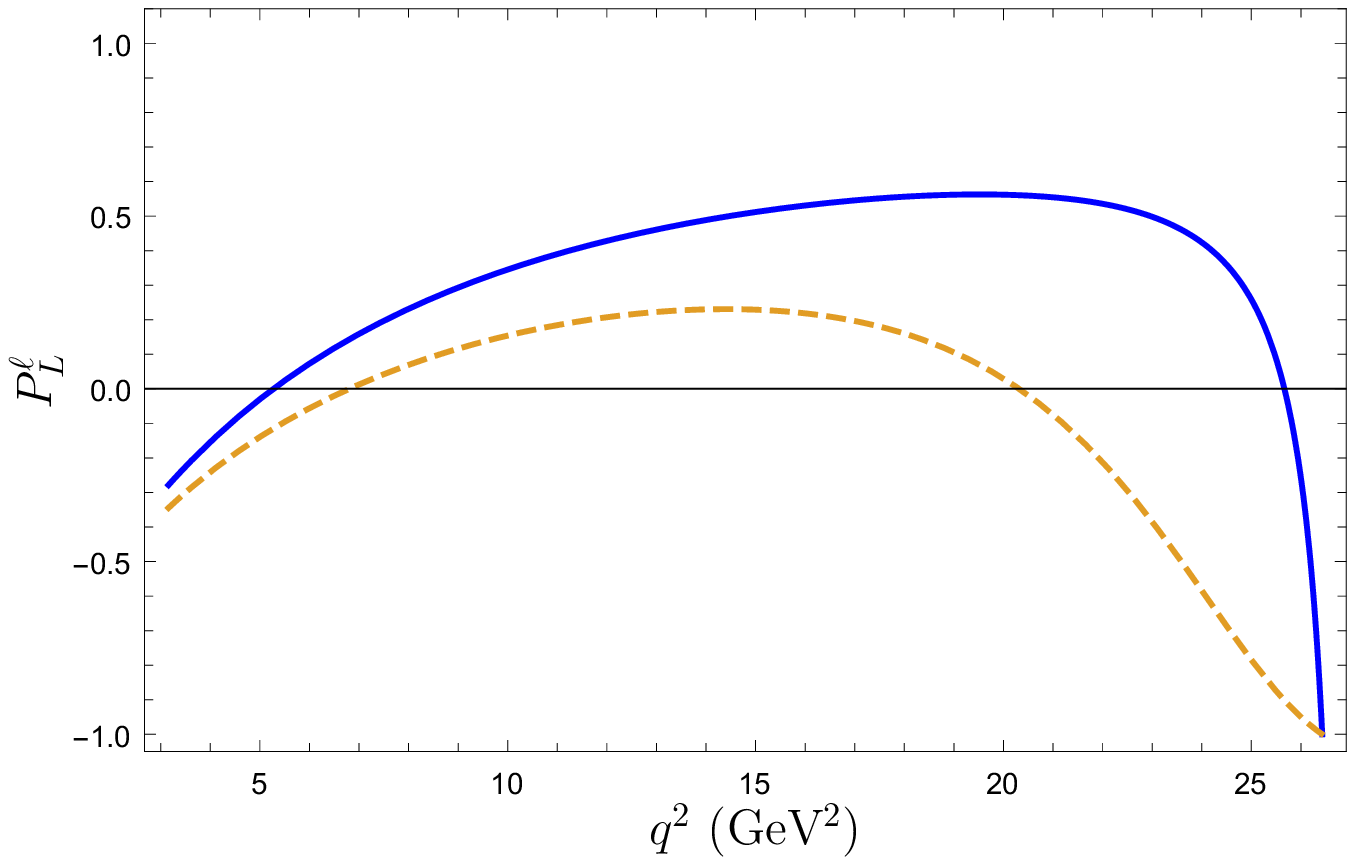}\ \
  \ \includegraphics[width=8cm]{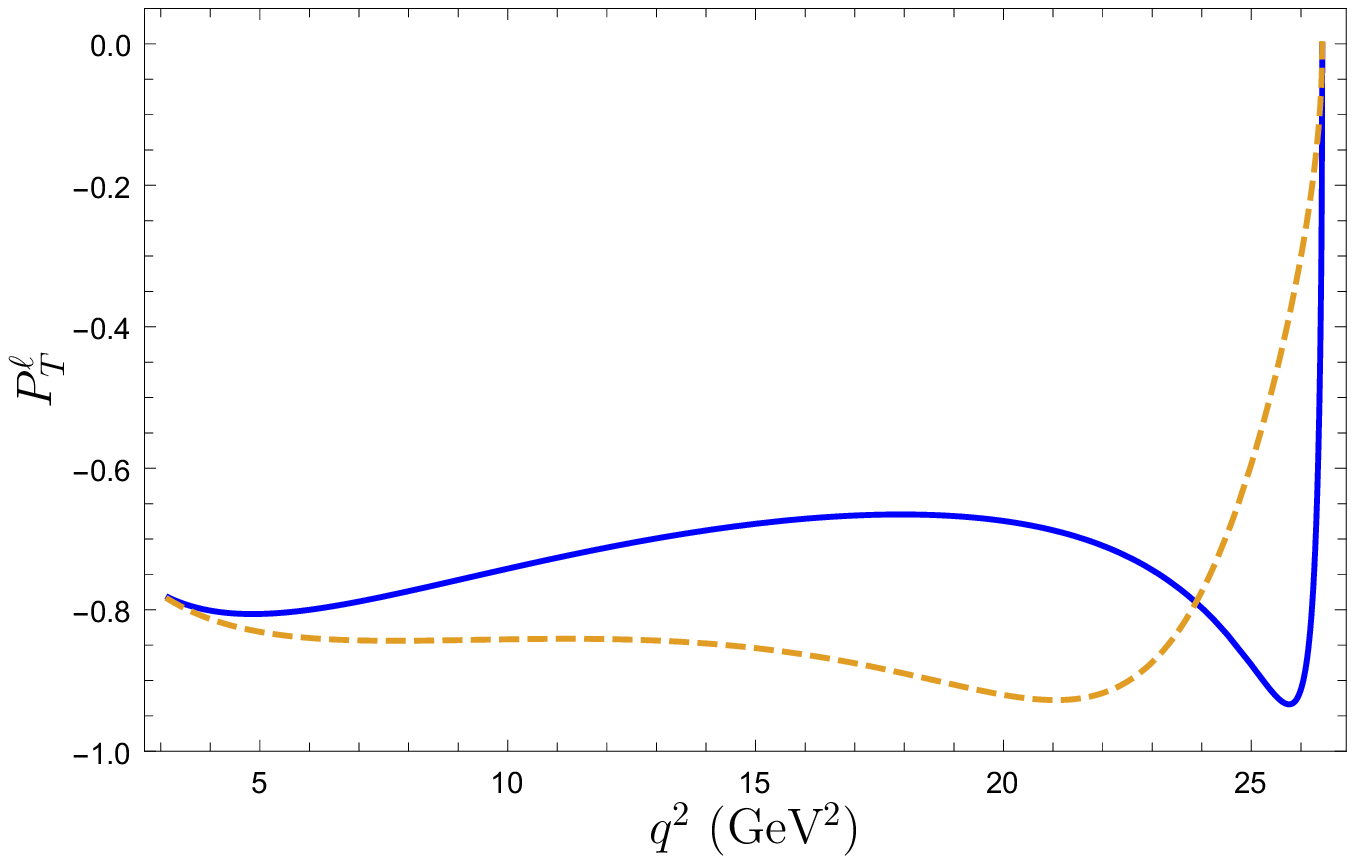}\\
\caption{Comparison of theoretical predictions for the differential
   FB asymmetry $A_{FB}$ and polarization $C^\ell_{F}$, $P^\ell_{L,T}$  parameters for the semileptonic
  $B\to \pi\tau^+\nu_\tau$ decay.  RQM result are given by  blue solid
  lines and  CLFQM results are given by orange dashed
  lines. }
\label{aspl}
\end{figure}

 In Fig.~\ref{aspl} we,
 as an example, present comparison of RQM and CLFQM predictions for
 the quantities $A_{FB}(q^2), C_F^\ell(q^2), P_{L,T}^\ell(q^2)$
 for the semileptonic  $B\to \pi\tau^+\nu_\tau$ decay. The
 RQM results are  plotted by the blue solid lines and CLFQM ones by
 orange dashed line. The $q^2$ dependence of these FB asymmetry and
 polarization parameters is determined by the corresponding dependence
 of the form factors and thus is significantly different.

\begin{table}
\caption{Comparison of RQM and CLFQM predictions with lattice data for
  FB asymmetry and lepton polarization for $B$ decays to
  light pseudoscalar mesons. }
\label{pikcomp}
\begin{ruledtabular}
\begin{tabular}{ccccccc}
Decay& \multicolumn{3}{c}{$\langle A_{FB}\rangle$} &
                                                       \multicolumn{3}{c}{$\langle
                                                       P^\ell_L\rangle$}
                \\  \cline{2-4}\cline{5-7}
   & RQM&CLFQM&Lattice \cite{msv}&RQM&CLFQM&Lattice \cite{msv}\\
  \hline
  $B\to \pi\mu^+\nu_\mu$ &$-0.004$&$-0.005$&$-0.0034(31)$  &0.99&0.98&0.988(9)\\
  $\bar B\to \pi\tau^+\nu_\tau$ &$-0.22$&$-0.28$&$-0.220(24)$  &0.42&0.087&0.301(86)\\
  $\bar B_s\to K\mu^+\nu_\mu$  &$-0.006$&$-0.007$&$-0.0046(28)$
              &0.98&0.98&0.986(7)\\
   $\bar B_s\to K\tau^+\nu_\tau$   &$-0.24$&$-0.29$&$-0.262(23)$  &0.35&$-0.10$&0.172(91)\\
\end{tabular}
\end{ruledtabular}
\end{table}

\begin{table}
\vspace*{-0.7cm}\caption{The branching ratios, FB asymmetry and polarization
  parameters of the semileptonic $B$ decays. For each channel, the result for RQM is shown in the upper line and the one for CLFQM in the lower line.}
\label{bcomp}
\begin{ruledtabular}
   \begin{tabular}{ccccccc}
Decay&Br&$\langle A_{FB}\rangle$  & $\langle C^\ell_{F}\rangle$& $\langle P^\ell_{L}\rangle$  &$\langle P^\ell_{T}\rangle$& $\langle F_{L}\rangle$ \\
\hline
\multirow{2}*{$B^+\rightarrow\bar D^0 e^+\nu_e$}&
$2.53\times10^{-2}$ & $-0.98\times10^{-6}$ & $-1.50$ & $1.00$ & $-1.01\times10^{-3}$ & \quad\\
&$2.23\times10^{-2}$ & $-1.04\times10^{-6}$ & $-1.50$ & $1.00$ & $-1.07\times10^{-3}$ & \quad\\
\hline
\multirow{2}*{$B^+\to \bar D^0\mu^+\nu_\mu$}&
$2.52\times10^{-2}$ &$-0.013$ & $-1.46$ & $0.96$ &$-0.19$ &\\
& $2.22\times10^{-2}$ & $-0.014$ & $-1.46$ & $0.96$ & $-0.20$ & \quad\\
\hline
\multirow{2}*{$B^+\to\bar D^0\tau^+\nu_\tau$}&
$0.68\times10^{-2}$&$-0.37$ & $-0.30$ & $-0.24$ &$-0.85$ &\\
&$0.67\times10^{-2}$&$-0.36$&$-0.27$&$-0.32$&$-0.84$&\\
\hline
\multirow{2}*{$B^+\rightarrow\bar D^{*0} e^+\nu_e$}&
$6.81\times10^{-2}$ & $-0.22$ & $-0.48$ & $1.00$ & $-0.34\times10^{-3}$ & 0.55\\
& $5.43\times10^{-2}$ & $-0.22$ & $-0.42$ & $1.00$ & $-0.29\times10^{-3}$ & 0.52\\
\hline
\multirow{2}*{$B^+\to\bar D^{*0}\mu^+\nu_\mu$}&
$6.77\times10^{-2}$  &$-0.23$ & $-0.47$ & $0.98$ &$-0.061$&0.55\\
& $5.70\times10^{-2}$ & $-0.23$ & $-0.41$ & $0.98$ & $-0.052$ & 0.52\\
\hline
\multirow{2}*{ $B^+\to\bar D^{*0}\tau^+\nu_\tau$}&
$1.56\times10^{-2}$  &$-0.32$ & $-0.060$ & $0.48$ &$-0.12$ &0.47\\
&$1.40\times10^{-2}$&$-0.30$&$-0.056$&$0.51$&$-0.10$&0.45\\
\hline
\multirow{2}*{$B^+\rightarrow\pi^0 e^+\nu_e$}&
$7.20\times10^{-5}$&$-0.28\times10^{-6}$&$-1.50$&$1.00$&$-0.46\times10^{-3}$&\\
&$7.89\times10^{-5}$&$-0.35\times10^{-6}$&$-1.50$&$1.00$&$-0.62\times10^{-3}$&\\
\hline
\multirow{2}*{$B^+\to \pi^0\mu^+\nu_\mu$}&
$0.72\times10^{-4}$&$-0.004$ & $-1.49$ & $0.99$ &$-0.09$ & \\
&$0.78\times10^{-4}$&$-0.005$&$-1.48$&$0.98$&$-0.12$&\\
\hline
\multirow{2}*{$B^+\to \pi^0\tau^+\nu_\tau$}&
$4.5\times10^{-5}$ &$-0.22$ & $-0.82$ & $0.42$ &$-0.72$ &\\
&$5.3\times10^{-5}$&$-0.28$&$-0.59$&$0.087$&$-0.85$&\\
\hline
\multirow{2}*{$B^+\rightarrow\rho^0 e^+\nu_e$}&
 $1.74\times10^{-4}$&$-0.50$&$0.042$&$1.00$&$-0.07\times10^{-3}$&0.31\\
& $2.19\times10^{-4}$&$-0.32$&$-0.39$&$1.00$&$-0.18\times10^{-3}$&0.51\\
\hline
\multirow{2}*{$B^+\to \rho^0\mu^+\nu_\mu$}&
$1.73\times10^{-4}$&$-0.51$ & $0.047$ & $0.99$ &$-0.011$&0.31\\
& $2.19\times10^{-4}$&$-0.32$&$-0.39$&$0.99$&$-0.03$&0.51\\
\hline
\multirow{2}*{$B^+\to \rho^0\tau^+\nu_\tau$}&
$0.97\times10^{-4}$&$-0.54$ & $0.14$ & $0.60$ &$0.095$ &0.31\\
&$1.19\times10^{-4}$&$-0.39$&$-0.12$&$0.60$&$-0.10$&$0.49$\\
\hline
\multirow{2}*{$B^+\rightarrow\eta e^+\nu_e$}&
$4.24\times10^{-5}$&$-0.37\times10^{-6}$&$-1.50$&$1.00$&$-0.60\times10^{-3}$&\\
&$5.44\times10^{-5}$&$-0.39\times10^{-6}$&$-1.50$&$1.00$&$-0.64\times10^{-3}$&\\
\hline
\multirow{2}*{$B^+\to \eta\mu^+\nu_\mu$}&
$0.42\times10^{-4}$&$-0.006$ & $-1.48$ & $0.98$ &$-0.12$ &\\
&$0.54\times10^{-4}$&$-0.006$&$-1.48$&$0.98$&$-0.12$&\\
\hline
\multirow{2}*{$B^+\to \eta\tau^+\nu_\tau$}&
$2.6\times10^{-5}$ &$-0.27$ & $-0.67$ &$0.21$ &$-0.83$ &\\
&$3.3\times10^{-5}$&$-0.29$&$-0.60$&$0.11$&$-0.86$&\\
\hline
\multirow{2}*{$B^+\to \eta'e^+\nu_e$}&
$3.17\times10^{-5}$&$-0.43\times10^{-6}$&$-1.50$&$1.00$&$-0.66\times10^{-3}$&\\
&$2.64\times10^{-5}$&$-0.49\times10^{-6}$&$-1.50$&$1.00$&$-0.72\times10^{-3}$&\\
\hline
\multirow{2}*{$B^+\to \eta'\mu^+\nu_\mu$}&
$0.31\times10^{-4}$&$-0.007$ & $-1.48$ & $0.98$ &$-0.13$ &\\
&$0.26\times10^{-4}$&$-0.007$&$-1.47$&$0.98$&$-0.14$&\\
\hline
\multirow{2}*{$B^+\to \eta'\tau^+\nu_\tau$}&
$1.7\times10^{-5}$ &$-0.30$ & $-0.59$ & $0.14$ &$-0.84$ &\\
&$1.4\times10^{-5}$&$-0.31$&$-0.52$&$0.026$&$-0.87$&\\
\hline
\multirow{2}*{$B^+\rightarrow\omega e^+\nu_e$}&
$1.71\times10^{-4}$&$-0.43$&$-0.17$&$1.00$&$-0.12\times10^{-3}$&0.41\\
&$2.08\times10^{-4}$&$-0.30$&$-0.42$&$1.00$&$-0.15\times10^{-3}$&0.51\\
\hline
\multirow{2}*{$B^+\to \omega\mu^+\nu_\mu$}&
$1.71\times10^{-4}$&$-0.43$ & $-0.16$ & $0.99$ &$-0.02$&0.41\\
&$2.07\times10^{-4}$&$-0.30$&$-0.41$&$0.99$&$-0.027$&0.52\\
\hline
\multirow{2}*{$B^+\to \omega\tau^+\nu_\tau$} &
$0.97\times10^{-4}$&$-0.49$ & $0.009$ & $0.59$ &$-0.007$ &0.40\\
     & $1.10\times10^{-4}$&$-0.36$&$-0.15$&0.65&$-0.06$&0.49\\
      \end{tabular}
\end{ruledtabular}
\end{table}

\begin{table}
\caption{The branching ratios, FB asymmetry and polarization
  parameters of the semileptonic $B_s$ decays. For each channel, the result for RQM is shown in the upper line and the one for CLFQM in the lower line.}
\label{bscomp}
\begin{ruledtabular}
   \begin{tabular}{ccccccc}
Decay&Br&$\langle A_{FB}\rangle$  & $\langle C^\ell_{F}\rangle$& $\langle P^\ell_{L}\rangle$  &$\langle P^\ell_{T}\rangle$& $\langle F_{L}\rangle$ \\
\hline
\multirow{2}*{$B_s\rightarrow D_s^- e^+\nu_e$}&
$2.12\times10^{-2}$&$-0.97\times10^{-6}$&$-1.50$&1.00&$-1.02\times10^{-3}$&\\
&$2.06\times10^{-2}$&$-1.05\times10^{-6}$&$-1.50$&1.00&$-1.07\times10^{-3}$&\\
\hline
\multirow{2}*{$B_s\to D_s^-\mu^+\nu_\mu$}&
$2.12\times10^{-2}$&$-0.013$ & $-1.46$ & $0.96$ &$-0.19$ &\\
&$2.05\times10^{-2}$&$-0.014$&$-1.46$&0.96&$-0.20$&\\
\hline
\multirow{2}*{$B_s\to D_s^-\tau^+\nu_\tau$}&
$0.61\times10^{-2}$&$-0.36$ & $-0.30$ & $-0.27$ &$-0.85$ &\\
&$0.61\times10^{-2}$&$-0.36$&$-0.26$&$-0.33$&$-0.84$&\\
\hline
\multirow{2}*{$B_s\rightarrow D_s^{*-} e^+\nu_e$}&
$5.06\times10^{-2}$&$-0.26$&$-0.35$&$1.00$&$-0.23\times10^{-3}$&$0.49$\\
&$5.07\times10^{-2}$&$-0.22$&$-0.43$&$1.00$&$-0.29\times10^{-3}$&$0.52$\\
\hline
\multirow{2}*{$B_s\to D_s^{*-}\mu^+\nu_\mu$}&
$5.05\times10^{-2}$   &$-0.27$ & $-0.33$ & $0.99$ &$-0.040$&0.49\\
&$5.05\times10^{-2}$&$-0.22$&$-0.41$&$0.98$&$-0.052$&$0.52$\\
\hline
\multirow{2}*{$B_s\to D_s^{*-}\tau^+\nu_\tau$}&
$1.23\times10^{-2}$ &$-0.32$ & $-0.040$ & $0.53$ &$-0.035$ &0.42\\
&$1.25\times10^{-2}$&$-0.29$&$-0.058$&$0.51$&$-0.10$&0.45\\
\hline
\multirow{2}*{$B_s\rightarrow K^- e^+\nu_e$}&
$15.6\times10^{-5}$&$-0.39\times10^{-6}$&$-1.50$&$1.00$&$-0.56\times10^{-3}$&\\
&$10.1\times10^{-5}$&$-0.43\times10^{-6}$&$-1.50$&$1.00$&$-0.72\times10^{-3}$&\\
\hline
\multirow{2}*{$B_s\to K^-\mu^+\nu_\mu$}&
$1.55\times10^{-4}$ &$-0.006$ & $-1.48$ & $0.98$ &$-0.11$ &\\
&$1.01\times10^{-4}$&$-0.007$&$-1.48$&$0.98$&$-0.14$&\\
\hline
\multirow{2}*{$B_s\to K^-\tau^+\nu_\tau$}&
$9.1\times10^{-5}$&$-0.24$ & $-0.77$ & $0.35$ &$-0.75$ &\\
&$6.8\times10^{-5}$&$-0.29$&$-0.46$&$-0.10$&$-0.86$&\\
\hline
\multirow{2}*{$B_s\rightarrow K^{*-} e^+\nu_e$}&
$3.29\times10^{-4}$&$-0.37$&$-0.23$&$1.00$&$-0.14\times10^{-3}$&0.44\\
&$3.30\times10^{-4}$&$-0.21$&$-0.59$&$1.00$&$-0.17\times10^{-3}$&0.59\\
\hline
\multirow{2}*{$B_s\to K^{*-}\mu^+\nu_\mu$}&
$3.29\times10^{-4}$   &$-0.38$ & $-0.22$ & $0.99$ &$-0.025$&0.44
    \\
&$3.29\times10^{-4}$&$-0.21$&$-0.59$&$0.99$&$-0.032$&0.59\\
\hline
\multirow{2}*{$B_s\to K^{*-}\tau^+\nu_\tau$}&
$1.82\times10^{-4}$  &$-0.44$ & $-0.032$ & $0.63$ &$-0.025$ &0.42\\
&$1.71\times10^{-4}$ & $-0.28$ & $-0.26$ & $0.65$ &$-0.13$ & $0.56$\\
 \end{tabular}
\end{ruledtabular}
\end{table}
Finally, we present a detailed
comparison of the branching ratios, FB asymmetry and polarization
parameters of the semileptonic $B$ and $B_s$ decays in RQM and
CLFQM  in Tables~\ref{bcomp} and \ref{bscomp}. The branching fractions are close for all decay modes. The
predictions for FB asymmetry and polarization parameters for the
heavy-to-heavy ($B\to D^{(*)}\ell\nu_\ell$ and
$B_s\to D_s^{(*)}\ell\nu_\ell$) semileptonic decays are also compatible in considered
models. This is expected, since the heavy-to-heavy form factors have similar $q^2$
behavior in both models. The main differences are found in these
parameters for the heavy-to-light
transitions. Therefore the
measurement of those observables can help to discriminate between
theoretical models.
\begin{itemize}
\item For decays to light pseudoscalar mesons, the most
sensitive observables are in the $\tau$ sector.

As was already noted (see Table~\ref{pikcomp}),
it is the longitudinal polarization fraction of $\tau$-lepton $\langle
P^\tau_L\rangle$, which values in models differ substantially and
even have the opposite signs for the $B_s\to K^-\tau^+\nu_\tau$ decay;
the other observable is the FB asymmetry $\langle A_{FB}\rangle$, which values are higher in
CLFQM. On the other hand, the absolute values of $\langle C^\tau_{F}\rangle$ are systematically higher in
RQM and for the $B_s\to K^-\tau^+\nu_\tau$ decay its value in RQM is
about 1.7 times larger than the one in CLFQM.

\item For decays to light vector mesons:

the main differences are in the lepton-side convexity parameter $\langle C^\ell_{F}\rangle$ which values differ significantly for all
decay modes and for $B^+\to \rho^0\ell^+\nu_\ell$ and $B^+\to \omega\tau^+\nu_\tau$
they even have different signs; the values of the FB asymmetry $\langle A_{FB}\rangle$ predicted by RQM are about a
factor of 1.5 higher than the ones of CLFQM for all decay modes, while
the values of the longitudinal polarization $\langle F_{L}\rangle$ in RQM are
about the same factor smaller than the ones in CLFQM.
\end{itemize}

\section{Conclusions}
\label{sec:concl}

The form factors of the semileptonic $B$, $B_s$ and $B_c$ decays are
calculated in the framework of the relativistic quark model (RQM) based on the quasipotential
approach in QCD. These form factors are expressed
trough the overlap integrals of the meson wave functions. The wave
functions of the initial and final mesons are taken from previous
calculations of the meson spectroscopy. All relativistic effects
including the
wave function transformations from the rest to moving reference frame
as well as contributions of the intermediate negative energy states are
consistently taken into account.  It is important to
emphasize that we have not used the heavy quark expansion for
description of the heavy-to-heavy meson decays and treated all decays
nonperturbatively in the inverse heavy quark mass.  Such approach allows us to obtain the
values of all form factors in the whole kinematical $q^2$ range
without extrapolations and additional model assumptions. The very
accurate and convenient analytic parameterizations, Eqs.~(\ref{fitfv}) and(\ref{fita12}),
of these form factors were obtained  with
parameters given in Table~\ref{mpf}.  These form factors
can be used for the calculations of various weak decays of bottom
mesons.

We use the calculated form factors for the evaluation of the
semileptonic branching fractions and differential distributions. In
particular we calculate mean values of the forward -backward asymmetry
$\langle A_{FB}\rangle$, lepton-side convexity parameter  $\langle
C^\ell_{F}\rangle$, longitudinal  $\langle P^\ell_{L}\rangle$ and
transverse $\langle P^\ell_{T}\rangle$ polarization of the charged
lepton, and the longitudinal polarization fraction $\langle F_{L}\rangle$ for the final-state vector meson.
These predictions are presented in Tables~\ref{asB}-\ref{asBc}.

Then we compare the results of our calculations with the covariant
light-front quark model (CLFQM). The comparison of the form factors
shows that both models give consistent values of all form factors
at the maximum recoil point of the final meson, $q^2=0$. The predicted values of the form
factors of the heavy-to-heavy  weak transitions at zero recoil point of the final meson,
$q^2=q^2_{\rm max}$, are also compatible. On the other hand, for  the
heavy-to-light weak transitions, where kinematical range is
substantially broader, the values of form factors at $q^2=q^2_{\rm max}$
differ significantly. The RQM form factors of these transitions grow
rapidly with $q^2$, while CLFQM form factors increase
moderately. Such difference in the form factor  $q^2$ dependence results
in significantly different differential distributions for bottom meson
decays to light mesons. We present a comprehensive comparison of RQM
and CLFQM predictions with results of lattice QCD
calculations and  available experimental data. It was found that the $q^2$
behavior of the form factor $f_+(q^2)$ for heavy-to-light $B\to \pi$
and $B_s\to K$ transitions as well as forward-backward asymmetry and polarization
parameters obtained in RQM agree better with lattice data than  CLFQM
ones. The differential distribution  $d Br(B\to\pi\mu\nu_\mu)/dq^2$ in
RQM is also in better agreement with averaged experimental data,
however the present experimental accuracy is not enough to distinguish between models. It
is found that some forward-backward asymmetry and polarization parameters
substantially differ in considered models and some of them have even
opposite signs. Such phenomenon also occurs frequently in nuclear physics --- results of spin observables can be quite different
in different models while the total cross sections agree rather well.
The most sensitive observables were identified. Thus their measurement can help to discriminate
between models and determine the $q^2$ dependence of the form factors. From a more recent and realistic perspective, one could measure
$A_{FB}(q^2)$ and $C_F^{\ell}(q^2)$ as a first step, which is feasible through an analysis to the two-dimensional $(q^2, \cos\theta)$ distribution
of event number with a large-statistics data. Actually we need to go beyond the measurement of an absolute branching fraction.

\acknowledgments
We are grateful to D. Ebert and M. Ivanov for valuable
discussions. The author X.W.K. acknowledges the support from the National Natural Science Foundation of China (NSFC) under
Project No. 11805012.

\end{document}